\documentclass[fleqn,usenatbib]{mnras}

\usepackage[T1]{fontenc}

\DeclareRobustCommand{\VAN}[3]{#2}
\let\VANthebibliography\thebibliography
\def\thebibliography{\DeclareRobustCommand{\VAN}[3]{##3}\VANthebibliography}

\usepackage[utf8]{inputenc}
\usepackage{subcaption}

\usepackage{graphicx}	
\usepackage{amsmath}	
\usepackage{amssymb}	

\usepackage{newtxtext,newtxmath} 

\title[Signatures OF AGN Feedback in NGC 5972]{Signatures of Feedback in the Spectacular Extended Emission Region of NGC 5972}

\author[T. Harvey et al.]{
Thomas Harvey,$^{1,2}$\thanks{E-mail: th4g18@soton.ac.uk}
W. Peter Maksym,$^{1}$
William Keel,$^{3}$
Michael Koss,$^{4, 5}$
Vardha N. Bennert,$^{6}$\newauthor
S. D. Chojnowski,$^{7}$
Ezequiel Treister,$^{8}$
Carolina Finlez,$^{8}$
Chris J. Lintott,$^{9}$
Alexei Moiseev,$^{10}$\newauthor
Brooke D. Simmons,$^{11}$
Lia F. Sartori,$^{12}$
and Megan Urry$^{13,14}$
\\
$^{1}$Center for Astrophysics | Harvard \& Smithsonian, 60 Garden Street, Cambridge, MA, 02138, USA\\
$^{2}$Physics and Astronomy, University of Southampton, Southampton, SO17 1BJ, UK\\
$^{3}$Department of Physics and Astronomy, University of Alabama, 206 Gallalee Hall, 514 University Blvd. Tuscaloosa, AL 35487-0324, USA \\
$^{4}$Space Science Institute, 4750 Walnut Street, Suite 205, Boulder, Colorado 80301\\
$^{5}$Eureka Scientific, 2452 Delmer Street Suite 100, Oakland, CA 94602-3017, USA\\
$^{6}$Physics Department, California Polytechnic State University, San Luis Obispo, CA 93407, USA\\
$^{7}$Department of Astronomy, New Mexico State University, Las Cruces, NM 88001, USA\\
$^{8}$Instituto de Astrof\'isica, Facultad de F\'isica, Pontificia Universidad Cat\'olica de Chile, Casilla 306, Santiago 22, Chile\\
$^{9}$Department of Physics, University of Oxford, Denys Wilkinson Building, Keble Road, Oxford OX1 3RH, UK\\
$^{10}$Special Astrophysical Observatory, Nizhny Arkhyz, 369167, Russia\\
$^{11}$Physics Department, Lancaster University, Lancaster LA1 4YB, UK\\
$^{12}$Institute for Particle Physics and Astrophysics, Department of Physics, ETH Zurich, Wolfgang-Pauli-Strasse 27, CH-8093 Zurich, Switzerland\\
$^{13}$Yale Center for Astronomy \& Astrophysics, Physics Department, P.O. Box 208120, New Haven, CT 06520, USA\\
$^{14}$Department of Physics, Yale University, P.O. Box 208121, New Haven, CT 06520, USA
}
\date{Accepted XXX. Received YYY; in original form ZZZ}

\pubyear{2022}

\begin{document}
\label{firstpage}
\pagerange{\pageref{firstpage}--\pageref{lastpage}}
\maketitle

\begin{abstract}
We present Chandra X-ray Observatory observations and Space Telescope Imaging Spectrograph spectra of NGC 5972, one of the 19 "Voorwerpjes" galaxies. This galaxy contains an Extended Emission Line Region (EELR) and an arc-second scale nuclear bubble. NGC 5972 is a faded AGN, with EELR luminosity suggesting a 2.1 dex decrease in L$_{\textrm{bol}}$ in the last $\sim5\times10^{4}$ yr. We investigate the role of AGN feedback in exciting the EELR and bubble given the long-term variability and potential accretion state changes.
We detect broadband (0.3-8 keV) nuclear X-ray emission coincident with the [O~{\sc iii}] bubble, as well as diffuse soft X-ray emission coincident with the EELR. The soft nuclear (0.5-1.5 keV) emission is spatially extended and the spectra are consistent with two APEC thermal populations ($\sim$0.80,$\sim$0.10 keV). We find a bubble age >2.2 Myr, suggesting formation before the current variability. We find evidence for efficient feedback with L$_{\textrm{kin}}/L_{\textrm{bol}}\sim0.8\%$, which may be overestimated given the recent L$_{\textrm{bol}}$ variation. Kinematics suggest an out-flowing 300  km s$^{-1}$ high-ionization [O~{\sc iii}]-emitting gas which may be the line of sight component of a $\sim$780 km s$^{-1}$ thermal X-ray outflow capable of driving strong shocks that could photoionize the precursor material. We explore possibilities to explain the overall jet, radio lobe and EELR misalignment including evidence for a double SMBH which could support a complex misaligned system.

\end{abstract}

\begin{keywords}
galaxies: Seyfert,
galaxies: active,
galaxies: ISM,
X-rays: galaxies,
individual: NGC 5972
\end{keywords}



\section{Introduction}

Active Galactic Nuclei (AGN) are thought to play an important role in the evolution of their host galaxy through their interaction with the interstellar medium via winds, outflows, and jets \citep[see reviews by][]{2012NewAR..56...93A, doi:10.1146/annurev-astro-081811-125521}. In many galaxies, direct photo-ionization by AGN emission only extends to the Narrow Line Region (NLR) within a kiloparsec of the supermassive black hole (SMBH) \citep[see e.g][]{bennert2002size, 2000ApJ...545...63E}. However there is a long known population of galaxies which display the same distinct ionization signatures on a much larger spatial scale; firstly surrounding radio-loud galaxies and quasars  \citep[see review by][]{2006NewAR..50..694S}, but more recently near lower luminosity AGN as well. These Extended Emission Line Regions (EELRs) contain AGN-ionized hot gas at scales larger than 10 kpc from the AGN \citep[see][]{Lintott2009, Schawinski2010, Keel2012, Keel2017}. These EELRs have been used as light echos to probe AGN variability on previously inaccessible kyr time-scales. 
\subsection{AGN Feedback}

AGN feedback is the idea that the interaction of outflows with the fuel supply of the AGN causes a feedback loop which self-regulates SMBH growth and star formation \citep{1998A&A...331L...1S, 2005gbha.conf..340D}. Infalling gas sets off a period of rapid, radiatively efficient, high Eddington ratio accretion until feedback from outflows entrains the infalling gas, choking the fuel supply and reducing the accretion rate and luminosity \citep[e.g.][]{2009ApJ...690...20S}. These duty cycles of high and low accretion, which we call AGN "state changes", may be analogous to observations of accretion in X-ray binaries (XRBs) where we observe rapid accretion rate variability on shorter time-scales of days to weeks \citep[see e.g.][]{2003MNRAS.345L..19M, mchardy2006active, 2006MNRAS.372.1366K, 2015MNRAS.450.3410E}. These X-ray binaries show both a luminous, radiatively efficient soft state and an radiatively inefficient hard state dominated by advection \citep{1995ApJ...452..710N} or jet accretion flow \citep{2004A&A...414..895F}. Each state corresponds to specific X-ray spectral properties (e.g. powerlaw slope) and the transition is thought to occur when the accretion rate is a few percent of the Eddington limit \citep{skipper2016probing, remillard2006x, belloni2010states}. For these XRBs \cite{2007A&ARv..15....1D} have shown that decreasing Eddington ratio is correlated with an increase in kinetic power. This idea has been extended to AGN \citep[see e.g.][]{2012NewAR..56...93A}, where luminous AGN correspond to the soft state and low-luminosity AGN correspond to the hard state, where most of the energy output is kinetic. Many authors  \citep{mchardy2006active, Schawinski2010}, have suggested that during this inefficient, kinetically dominated accretion state the SMBHs perform significant feedback work on the interstellar medium (ISM) within their host galaxies. The periods of rapid efficient accretion are thought to last for around 10$^5$ years, with fast state changes on the order of 10$^{4}$ years \citep{Keel2012,Keel2017, 10.1093/mnras/stv1136}. The AGN flicker on and off for an overall total accretion time-scale of 10$^{7-9}$ years  \citep[e.g.][]{1982MNRAS.200..115S, 2002MNRAS.335..965Y}. 

The mechanisms and effects of this feedback on the wider galaxy are not well understood. We have known for some time that the masses of SMBH are correlated with properties of their host galaxies  \citep[e.g.][]{1998AJ....115.2285M, 2000ApJ...539L..13G, 2007ApJ...669...67H} and it is thought that this feedback interaction leads to these correlations \citep{1998A&A...331L...1S, 10.1046/j.1365-8711.1999.03017.x}. 

\cite{10.1111/j.1365-2966.2009.15643.x} have suggested a two-stage model for AGN feedback, where small high velocity outflows interacting with the diffuse warm/hot ISM cause additional feedback effects that lead to the dissipation of the cold ISM on large scales. The cold ISM is primarily composed of dense molecular clouds, which are the primary reservoir for star formation \citep[e.g.][]{mckee1977theory}. This idea reduces the kinetic output of the AGN required to unbind the ISM (and halt star formation) by an order of magnitude compared to conventional feedback models (\citep[e.g.][]{2009AIPC.1201...33M}. \cite{10.1111/j.1365-2966.2009.15643.x} estimates around 0.5\% of $L_{\textrm{bol}}$ is required for efficient feedback under this two-stage model.

\subsection{Voorwerpjes}
Hanny's Voorwerp (HV), discovered by Hanny von Arkel through the Galaxy Zoo citizen science initiative \citep{Lintott2009} is one of the first known EELRs associated with a faded AGN. HV is an ionized cloud $\approx$20 kpc from IC 2497, which appears blue in SDSS \textit{gri} images due to strong emission in [O~{\sc iii}]$\lambda4960, 5008$\AA. HV is also the first example of a quasar "light echo". Analysis by \cite{Lintott2009} and \cite{Keel2012hst} estimate the luminosity required to sustain the EELR emission based on ionization parameter and recombination arguments as 1-4 $\times 10^{45}$ erg s$^{-1}$. This energy imbalance suggests the luminosity required to sustain the observed EELR emission is 50 times greater than the current bolometric luminosity based on \textit{Wide-field Infrared Survey Explorer (WISE)} mid-IR observations. X-ray observations by \cite{10.1093/mnras/stx2952} show the nucleus is heavily obscured (\textit{N$_{\textrm{H}}$}$ \approx 2 \times 10^{24}$ cm$^{-2}$), so the prevailing hypothesis is that the intrinsic luminosity of IC 2497 has decreased by $50\times$ in the last 100 kyr, but we still see emission from HV due to the light travel time and recombination time-scale of the narrow-line emission \citep{Keel2012hst}. \cite{sartori2018model} finds that the luminosity drop  in the last 100 kyr corresponds to a decrease in Eddington ratio from 0.35 to 0.007. The comprehensive study of HV has demonstrated the usefulness of EELRs as probes of AGN variability on time-scales of 10$^{4-5}$ years \citep{Lintott2009,Schawinski2010,Sartori2016,sartori2018model,Keel2012,Fabbiano2019}. The information encoded within EELRs allows us to probe the variability of these sources on the time-scales at which we see significant changes \citep{10.1093/mnras/stv1136}.  

Following the discovery of HV, Galaxy Zoo volunteers performed a search of archival SDSS observations looking for objects with similar colours and morphologies. \cite{Keel2012} found 19 candidates with EELR >10kpc from the SMBH. 7 of these were found to have a luminosity deficit similar to HV and in several the energy mismatch is significant enough that they may also be examples of faded AGN. As noted by \cite{Keel2017}, most of the known EELR systems are post-merger or interacting systems which contain the significant regions of extraplanar cold gas required to produce EELRs. \cite{2018MNRAS.476L..34S,2019ApJ...883..139S} present a framework to link observed AGN variation of different magnitudes and timescales, such as the accretion stage changes suggested by the "Voorwerpjes". 
\subsection{NGC 5972}
One of the 19 galaxies identified as hosting an EELR is NGC 5972, due to its distinctive double-helix shaped emission line regions. \cite{Keel2012} identified it as a faded AGN because the ratio of cloud-ionising to bolometric luminosity was calculated to be >1.8. NGC 5972 hosts a powerful (\textit{L} $\approx 10^{44} \textrm{ erg s}^{-1}$), highly obscured (\textit{log N$_{\textrm{H}}$}=24.34, \cite{Zhao2020}) AGN which powers a spectacular double helix-shaped [O~{\sc iii}] emission structure. The EELR arms were originally identified as ionised gas and not a spiral galaxy by \cite{1995A&A...296..315V}. NGC 5972 is a nearby (\textit{z}=0.02964, \textit{d}=133$\pm$9 Mpc, \cite{2014MNRAS.440..696A}) Seyfert 2 galaxy which is thought to have faded dramatically in the last 50,000 years. It is radio-loud (1.4 Ghz Luminosity $\approx$ 2 $\times 10^{24}$ W Hz$^{-1}$, \cite{condon1998nrao}), with a classic double-lobed radio structure separated by 330 kpc \citep{Keel2012}. \cite{Keel2012} note that the misalignment of the EELR and radio lobes by 67$^\circ$ is unusual for Seyfert galaxies and suggest that either the ionization cones have large opening angles, or the radio jets twist close to the AGN. 

\cite{Keel2015} used Hubble Space Telescope (HST) imaging and supporting ground-based observations to investigate the morphology of NGC 5972 at sub-arcsecond resolution. They find evidence of a past merger via tidal tails and dust lanes consistent with a  a 1.5 Gyr-old precessing warped disc. The outermost parts of the EELR are located up to  50 kpc from the nucleus and follow the main disk  rotation curve. In \cite{Keel2017} the authors present a detailed analysis of the luminosity history of NGC 5972 over the last 55 kyr. They find evidence for a 2.1 dex decrease in bolometric luminosity over this time period.

\cite{Keel2017} draw attention to an arc-second scale [O~{\sc iii}] structure, which hosts the central engine. They hypothesise that this feature, which doesn't show any velocity structure, is a loop or bubble of hot ionized gas. Line ratios suggest there are lower abundances within the bubble. The authors hypothesise that the bubble is actually a signature of infalling gas rather than an AGN-driven outflow. Figure \ref{fig:stis_loc} reproduces their [O~{\sc iii}] image of this structure which we hereafter refer to as the "[O~{\sc iii}] bubble" or just the "bubble". 

In \cite{Zhao2020}, analysis of \textit{Nuclear Spectroscopic Telescope Array (NuSTAR)} and \textit{Chandra X-Ray Observatory} observations was used to derive properties of the AGN and torus within NGC 5972. The authors extracted a spectra from the nucleus and fitted a physically-motivated AGN model which constrains parameters such as the column density (both in the LOS and averaged across the torus), torus angle, covering factor, and Powerlaw Index. They find NGC 5972 to host a highly obscured, Compton-thick AGN (log N$_{\textrm{H, tor}}=24.34$) which we observe through the obscuring torus. 

We aim to understand the interaction and feedback effects between the hot gas in the bubble and the SMBH given the AGN state changes over the past $10^5$ years. In this paper we present new analysis of high resolution Chandra X-ray imaging and spectroscopy of the AGN and EELR in NGC 5972. Chandra's sub-arcsecond resolution allows us to study the interaction of the central engine and surrounding gas on small scales. 
We also use new HST Space Telescope Imaging Spectrograph (STIS) spectroscopy to investigate the kinematics of the gas within the [O~{\sc iii}] bubble. Together these observations of the velocity, temperature and ionization source of gas within the [O~{\sc iii}] bubble allow insights into the relationship between the nuclear SMBH and its environment.

This paper is organised as follows. In Section \ref{sec:obs_data_reduc} the observations and data reduction process are described. In Section \ref{sec:analysis} we detail our methodology. The results are presented and discussed in Sections \ref{sec:results} and \ref{sec:discussion} respectively. Finally we summarise our findings in Section \ref{sec:conclusions}. 

For distances and angular sizes we adopt the cosmological parameters \textit{$H_0$} = 70 km s$^{-1}$ Mpc$^{-1}$, \ \textit{$\Omega_{m, 0}$} = 0.3, \textit{$\Omega_{\Lambda, 0}$} =0.7. Emission line wavelengths are given in vacuum and taken from the \textit{Sloan Digital Sky Survey}\footnote{\url{http://classic.sdss.org/dr6/algorithms/linestable.html}}.

\section{Observations and Data Reduction}
\label{sec:obs_data_reduc}
\subsection{X-ray Observations}
Three observations of NGC 5972, totalling 55 ks, were obtained from the Chandra Data Archive\footnote{\url{http://cda.harvard.edu/chaser}}. These observations were taken using the Advanced CCD Imaging Spectrometer (ACIS), centered on the back-illuminated S3 chip within the ACIS-S array. The dates and lengths of these observations are given in Table \ref{tab:chandra_obs}. The observations were reduced using the Chandra Interactive Analysis of Observations (CIAO) package \citep{2006SPIE.6270E..1VF} version 4.1.3 and version 4.6.8 of the Chandra Calibration DataBase (CALDB). The observations were reprocessed using the standard \textit{chandra\_reprocess} script, followed by deflaring (using \textit{deflare}) to remove periods of high background at a confidence level of 3$\sigma$. NGC 5972 itself was excluded from the deflaring process. There were very few periods of high background, and the lengths of the total and useful observations are given in Table \ref{tab:chandra_obs}. 

As is now the default option in CIAO, the observations were reprocessed to take advantage of the sub-pixel capabilities of the ACIS detector due the spacecraft dither. The Energy Dependent Subpixel Event Repositioning algorithm \citep[EDSER][]{li2004chandra} can improve the spatial binning of the inferred event positions, increasing spatial resolution from the native pixel size of 0$\farcs$492 to 0$\farcs$1-0$\farcs$2. 

The individual observations were astrometrically aligned to the longest observation (obsid 19562). Firstly the \textit{fluximage} tool was used to generate an image with 1/2 pixel binning (0$\farcs$256) from the event files, then a source list was extracted using \textit{wavdetect}. The source lists were compared using \textit{wcs\_match} and the astrometry of the event and aspect files were corrected using \textit{wcs\_update}. We allow only translational alignment of the images to avoid calibration issues with the aligned data products. 

In order to identify as many sources as possible to obtain the best match we combine the observations using \textit{merge\_obs} and use the same process as above to align the merged event file to the Pan-STARRS DR1 catalog \citep{2016arXiv161205560C}. The final reported Root Mean Square offset between the extracted source list and the catalog is 0$\farcs$09. This ensures we have a precise astrometric match to the HST imaging to allow comparison of small-scale features. The computed transform was then used to align each individual observation with the Pan-STARRS catalog. 
\begin{table*}
\centering
\caption{Details of the Chandra ACIS-S observations used in this analysis.}
\begin{tabular}{lllll}
Observation ID & Date       & Total Exposure (ks) & Useful Exposure (ks) & PI      \\ 
\hline
18080          & 2016-04-04 & 9.83          & 9.83                    & Ajello         \\
19562          & 2017-12-18 & 23.74         & 23.6                    & Maksym            \\
20893          & 2017-12-19 & 23.73         & 23.2                    & Maksym            \\ 
\hline
Merged         &            & 57.3          & 56.6                    &                \\
\hline
\end{tabular}
\label{tab:chandra_obs}
\end{table*}

\subsection{Optical Data}

\subsubsection{HST Imaging}
HST imaging of NGC 5972 has been analysed extensively by \cite{Keel2015} and \cite{Keel2017}. They obtained emission line images of the [O~{\sc iii}]$\lambda$5008\AA \ line in the Advanced Camera for Surveys (ACS) FR505N ramp filter and H$\alpha$+[N~{\sc ii}] images with Wide Field Camera 3 (WFC3) F673N narrowband filter. Details of these observations are given in Table \ref{tab:hst_obs}. We obtained these observations from the Hubble Legacy Archive and reduced them following the steps in \cite{Keel2015} to produce continuum subtracted and astrometrically matched images of NGC 5972 in flux units of photon cm$^{-2}$ s$^{-1}$. 

\begin{table}
\centering
\caption{Details of the Hubble Space Telescope observations used in this paper.}
\begin{tabular}{llll}
Filter       & Date       & Exposure (s) & PI      \\ 
\hline
Imaging      &            &              &         \\ 
\hline
ACS FR505N   & 2012-07-01 & 2538         & Keel    \\
WFC3 F673N   & 2012-12-18 & 2890         & Keel    \\ 
\hline
Spectroscopy &            &              &         \\ 
\hline
STIS G430L   & 2016-07-09 & 2071.8       & Maksym  \\
STIS G750M   & 2016-07-09 & 2854.8       & Maksym  \\
\hline
\end{tabular}
\label{tab:hst_obs}
\end{table}
\subsubsection{HST Spectroscopy}
\label{sec:hst_spec_reduce}
\begin{figure}
	\includegraphics[width=\columnwidth]{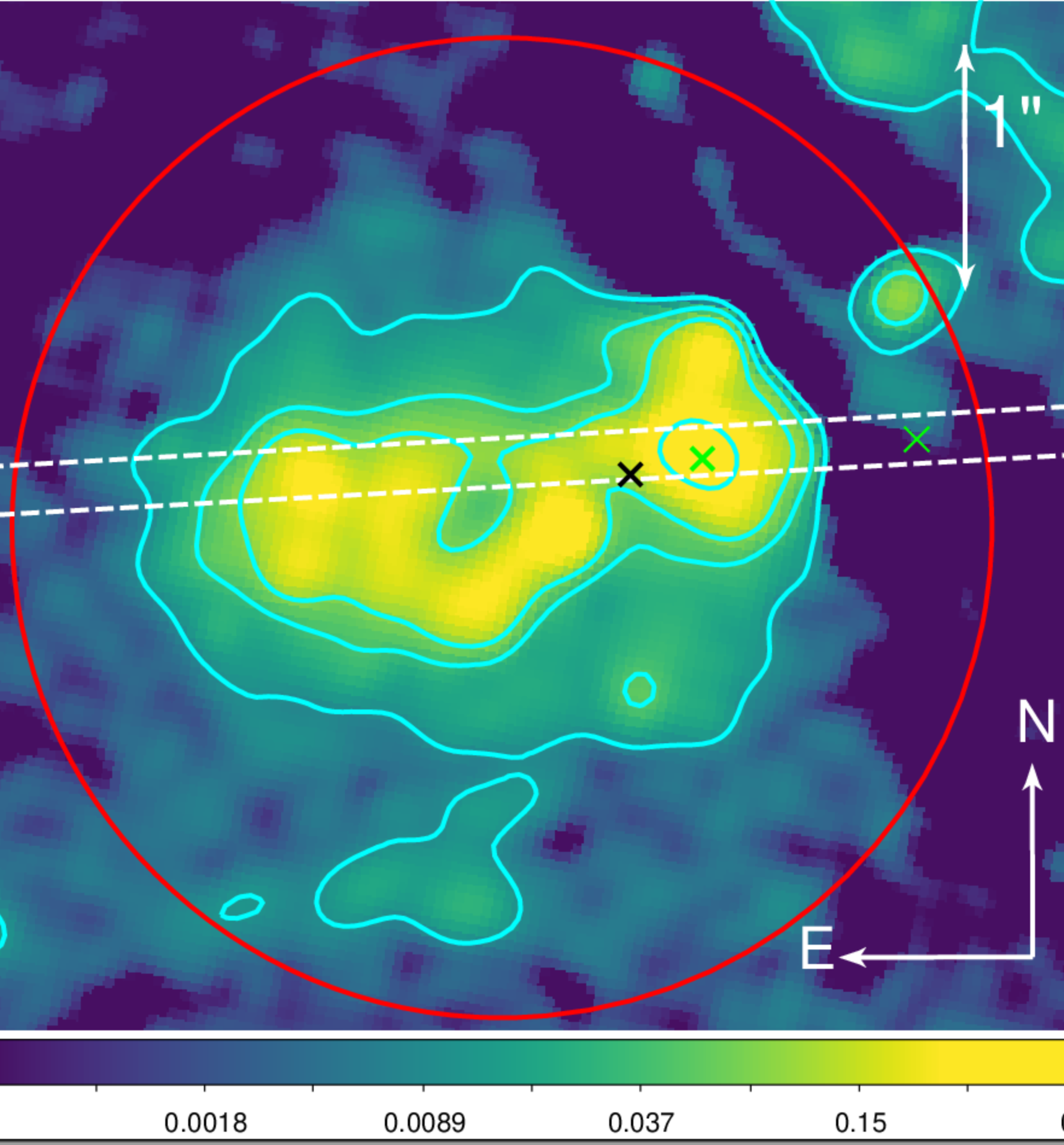}
    \caption{The STIS slit location for the observations detailed in Section \ref{sec:hst_spec_reduce}, overlaid on the HST ACS WFC3 narrowband [O~{\sc iii}] image with a log scale. The inferred nuclear source location (detailed in Section \ref{sec:agn_loc}) and the brightest [O~{\sc iii}] region is shown by the central green cross. Logarithmic contours, with 5 contour levels and a dynamic range of 0.0015 to 0.01 photon s$^{-1}$ cm$^{-2}$ pixel$^{-1}$, are shown in blue. The red circle (2$\arcsec$ radius) is the same as shown in other figures in the Results section. }
    \label{fig:stis_loc}
\end{figure}

\begin{figure}
    \centering
    \includegraphics[width=\columnwidth]{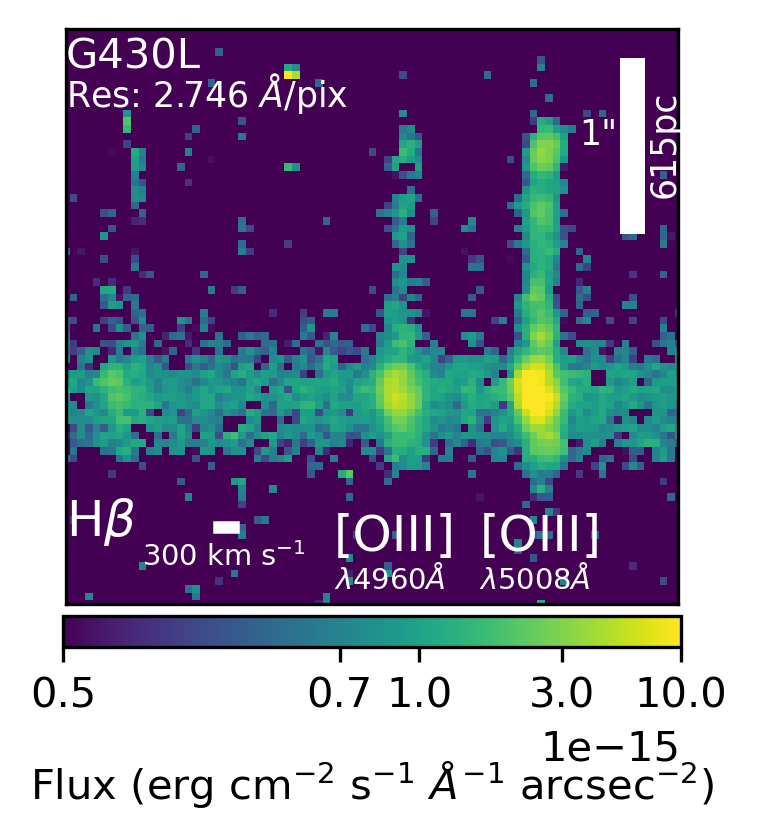}
    \caption{2D spatially resolved spectra of the nuclear bubble showing the [O~{\sc iii}]5008\AA \ and 4960\AA \ emission line profiles from the HST G430L observation. The horizontal (wavelength) scale is shown in velocity units, and the vertical (spatial) scale is indicated in arcseconds and parsec. A logarithmically scaled colorbar indicates the flux of each pixel.}
    \label{fig:g430l_2d}
\end{figure}

\begin{figure}
    \centering
    \includegraphics[width=\columnwidth]{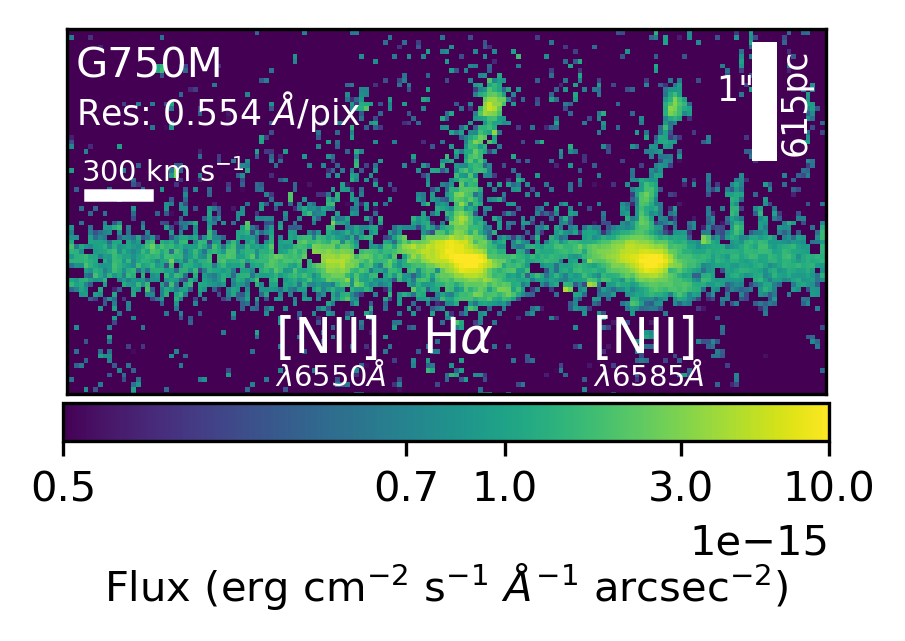}
    \caption{2D spatially resolved spectra of the nuclear bubble showing the H$\alpha$, [N~{\sc ii}]6550\AA and 6585\AA emission line profiles from the HST G750M. observation. The horizontal (wavelength) scale is shown in velocity units, and the vertical (spatial) scale is indicated in arcseconds and parsec. A logarithmically scaled colorbar indicates the flux of each pixel.}
    \label{fig:g750m_2d}
\end{figure}

Slit spectroscopy of NGC 5972 was taken during HST Cycle 23 (PI Maksym, Proposal ID 14271). The sightline used was positioned to obtain spectra from the arc-second scale [O~{\sc iii}] bubble seen in the HST [O~{\sc iii}] imaging. The slit location is shown in Figure \ref{fig:stis_loc}, with the slit centre located at RA. 15:38:54.0760 decl +17:01:34.590 (shown by green cross on far right) at a position angle (P.A) of 93$^\circ$. The observations were taken using the Space Telescope Imaging Spectrograph (STIS) with gratings G430L and G750M. The slit used has an aperture of 52$\arcsec$x0$\farcs$2. Details of the observations are given in Table \ref{tab:hst_obs}. Each observation was split into 3 components for cosmic ray rejection, but there is still a high cosmic ray background in the processed 2D spectra produced by the Hubble Legacy Archive's automated pipeline. To reduce this we obtained the raw data files and used the Python \textit{astroscrappy} \citep{curtis_mccully_2018_1482019} implementation of the L.A Cosmic algorithm \citep{2001PASP..113.1420V} to remove cosmic rays. The tool was run with a threshold value of 5$\sigma$ which removed most of the cosmic rays. The raw files were then processed using the \textit{calstis} script from the Python package \textit{stistools\footnote{\url{https://stistools.readthedocs.io/en/latest/}}} to produce cosmic ray rejected 2D spectral images for each filter. Reprocessing the data also decreases the systematic uncertainties in the astrometry to <0\farcs01 due to the addition of Gaia data into the pipeline\footnote{\url{https://outerspace.stsci.edu/display/HAdP/Improvements+in+HST+Astrometry}}.
Figure \ref{fig:g430l_2d} and Figure \ref{fig:g750m_2d} show the 2D spatially resolved spectra for both filters, where the horizontal axis corresponds to wavelength and the vertical axis is spatial offset. The spectral regions corresponding to bright emission lines are labelled, and a scale for both axis is shown. 
\subsection{Radio Data}

\begin{figure*}
    \centering
    \includegraphics[width=\textwidth]{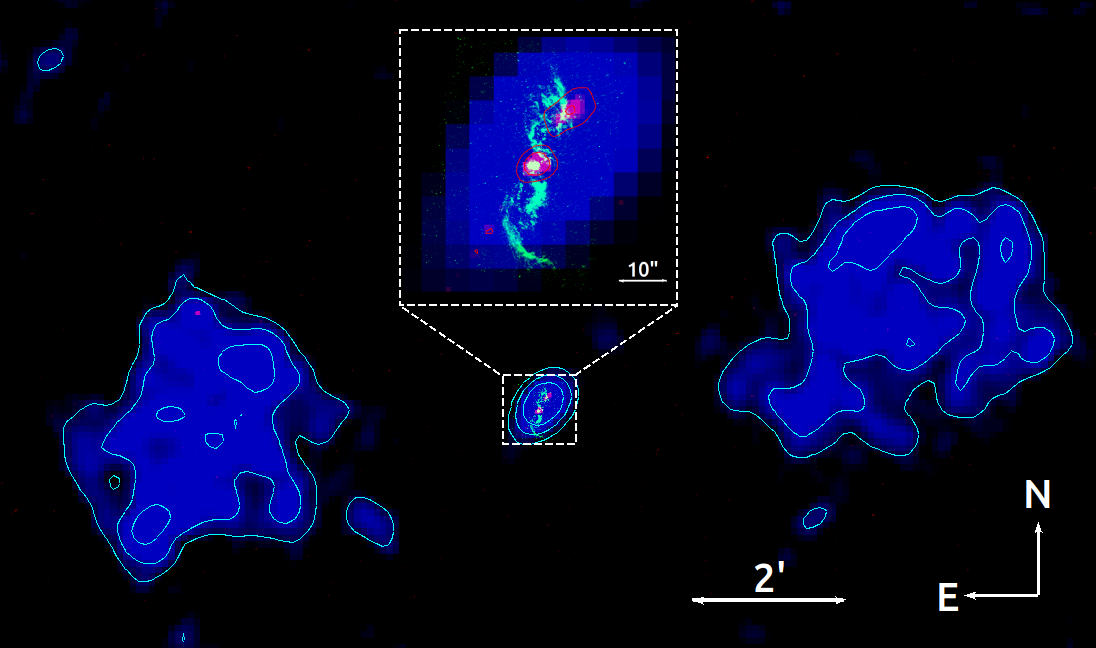}
    \caption{A three colour image of the radio and [O~{\sc iii}] emission from NGC 5972. The blue and red channels show VLA observations at different wavelengths (VLASS 10 cm and VLA 20 cm respectively) which are highlighted by logarithmic contours of the corresponding colour. The blue radio lobe contours have 3 levels with a dynamic range of 0.0007 to 0.003 Jy/beam, and the red contour shows one level of 0.0003 Jy/beam to indicate strong radio emission. The HST [O~{\sc iii}] is shown in green for reference, and the cutout shows an enlarged view of the [O~{\sc iii}] structure overlaid with the jet emission in red.}
    \label{fig:radio_lobes}
\end{figure*}

As NGC 5972 is radio-loud we obtained archival Very Large Array (VLA) observations in order to make comparisons between the X-ray, optical and radio morphologies.  A recent observation of the field containing NGC 5972 from the Very Large Array All Sky Survey was downloaded \citep[VLASS][]{2020PASP..132c5001L} using the CIRADA Cutout service\footnote{\url{https://ws.cadc-ccda.hia-iha.nrc-cnrc.gc.ca/en/doc/data/}}. It was taken at a central wavelength of 10 cm on 2019-03-21 and has a resolution of 2$\farcs$5.  We also obtain an archival 20cm VLA observation from the National Radio Astronomy Observatory Archive\footnote{\url{https://data.nrao.edu/portal/}} taken on 1986-11-26. For both observations the default pipeline processing is suitable for our purposes as we are only interested in the bright, large scale morphology. Figure \ref{fig:radio_lobes} shows the 20 cm observation in blue, the VLASS 10 cm observation in red and the HST [O~{\sc iii}] image of NGC  5972 in green. The central cutout shows an enlarged view of the small-scale radio jet overlaid on the EELR features. The red (cyan) contours indicate the regions of strong radio emission in the 10 cm (20 cm) observation. Purple contours in Figure \ref{fig:arms_oiii_xray_smooth} also show the jet emission observed at 10 cm (3 GHz) with VLASS.  

\subsection{Ground Based Observations}

We obtained near-infrared (NIR) observations of NGC 5972 in order to constrain the separation of a potential binary SMBH system.
NGC 5972 was observed in the NIR with laser guide star adaptive optics on the Keck 2 telescope at the W. M. Keck Observatory with the Near-Infrared Camera 2 (NIRC2) on 2022-05-13 as part of program study Swift BAT AGN (PI M. Urry). The observations used the wide camera, which has a 40 arcsec field of view (FOV) and 40 milli-arcsecond pixel s$^{-1}$ following the survey strategy and reductions in \cite{koss2018population}. We used a three-point dither pattern for 18 min in the Kp filter (2.1 \micron, bandpass 0.35 \micron). The images were reduced and combined with the Keck AO Imaging (KAI) data reduction pipeline \citep{jessica_lu_2022}. The image is astrometrically aligned using GAIA-detected stars present in the image.

\section{Method}
\label{sec:analysis}
\subsection{Chandra Imaging}

We produced 1/4 pixel (0$\farcs$124) resolution images of NGC 5972 in the soft X-ray (0.3 - 2 keV), hard X-ray (2-8 keV), Fe K$\alpha$ line (6-7 keV) and Ne IX line (0.865-0.938 keV). We have used the \textit{dmimgadapt} algorithm to produce adaptively smoothed images where large scale features are smoothed while  those on small scales  are preserved. We produce adaptively smoothed images at both 3$\sigma$ (9 counts) and 1$\sigma$ (3 counts) to look for structure within the extended soft X-ray emission at different levels of significance.

In order to determine the nuclear source location we look for the telltale strong Iron K$\alpha$ line which is evidence of emission from the outer accretion disk and inner torus \citep{reynolds2012constraints,gandhi2015dust}. This hard emission is expected to be a point-like source subject to Chandra's Point Spread Function (PSF). The neutral iron line and other nuclear iron lines, (e.g. Fe~{\sc xxv}, Fe~{\sc xxvi}) emit between 6 and 7 keV. We limited the event file to this energy range and produced an image binned to 1/4 ACIS pixel resolution using \textit{dmcopy}. Our initial method was then to use the \textit{wavdetect} to do source detection, generating an elliptical region centered on the peak of Fe K$\alpha$ emission. We found issues with reproducibility leading to a modified method discussed in Section \ref{sec:agn_loc}. 

\subsection{Chandra Spectroscopy}
\subsubsection{Spectral Extraction}
In order to investigate the emission from the nucleus (coincident with the [O~{\sc iii}] bubble) we extracted spectra from both a 0$\farcs$25 and 2$\arcsec$ radius annuli centered on the nuclear source location. The nuclear source location used is discussed in Section \ref{sec:agn_loc}. A 6$\arcsec$ radius region 55$\arcsec$ E of the nucleus was used to estimate the background. The HST [O~{\sc iii}] image was used to define elliptical annuli covering the north and south EELRs. Figure \ref{fig:arms_oiii_xray_smooth} shows these regions with dashed white ellipses. 
Spectra were extracted from these regions using the Ciao tool \textit{specextract} following standard procedures and analysed using the Python package Sherpa \citep{2007ASPC..376..543D}. Spectra were binned to 5 counts/bin to maximise the spectral resolution. The regions contained relatively few counts, between 15 and 300 depending on region and observation. For less than 25 counts per bin the $\chi^2$ statistic is unsuitable, and instead we use the \textit{wstat} statistic from Xspec. This statistic assumes that both the data and background obey a Poisson distribution and includes the background within the fit without modelling it. The same statistic and the Sherpa tool \textit{conf} are used to estimate the $1\sigma$ uncertainties in the best-fitting parameters.  

\subsubsection{Spectral Modelling of Nucleus}
\label{sec:nucleus_spec}
The medium-hard X-ray emission from the nucleus of NGC 5972 has been modelled by \cite{Zhao2020} based on the archival Chandra X-ray data (1-7 keV) as well as\textit{ Nuclear Spectroscopic Telescope Array} (NuSTAR) data (3-78 keV). As NGC 5972 is Compton-thick, the harder energy range covered by NuSTAR is needed to constrain AGN properties such as the powerlaw slope. Extended emission from hot gas in the bubble and NLR is contaminated with emission from the central engine due to PSF spread. We have to constrain the emission from the AGN itself before we introduce models for the extended soft emission.

\cite{Zhao2020}'s Xspec model consists of a physically motivated clumpy torus model derived from Monte Carlo simulations \citep[borus02][]{2019RNAAS...3..173B} plus an absorbed cutoff powerlaw (\textit{plcutoff}$\times$\textit{phabs}$\times$\textit{cabs}) and a small fractional constant representing unabsorbed reflected emission (\textit{const}$\times$\textit{plcutoff}). The entire model is multiplied by a \textit{phabs} absorption component representing galactic absorption, with N$_H$ =$3.12\times 10^{20}$ cm$^{-2}$ from the Colden Calculator \citep[NRAO dataset][]{dickey1990hi} in the CXC Proposal Planning Toolkit. They fit the X-ray emission from NGC 5972 using an extraction region of radius 5$\arcsec$ (75$\arcsec$) for Chandra (NuSTAR) observations. We freeze their best-fitting values $\Gamma = 1.75$, log N$_{\textrm{H, los}} = 23.79$, log N$_{\textrm{H, tor}} = 23.34$, cos($\theta_{\textrm{inc}}$) = $0.856$ and c$_{f} = 0.9$ in our model. Firstly we fit the hard emission (2-8 keV) within 0$\farcs$25 of the central engine to fix the fractional constant which represents reflected emission and dominates between 2 and 4 keV. We then fit the spectra from the 2$\arcsec$ bubble region for the same energy range to constrain the overall scaling factor of the AGN model, before introducing further model components to model the soft emission below 2keV.

To characterise the soft emission we explored combinations of the physically motivated models used for Mrk 573 by \cite{Paggi2012}. This includes a component representing collisionally ionized diffuse gas  \citep[APEC,][]{2001ApJ...556L..91S} and the CLOUDY derived \citep{ferland1998cloudy} table model used by \cite{Paggi2012}, which represents emission from gas photoionized by an AGN spectra. The CLOUDY model fits a grid of ionization parameters ($-2 \leq \textrm{log U} \leq 3$) and column densities ($19 \leq \textrm{log N}_{\textrm{H}} \leq 24$). 

We fit the 2 longest and concurrent observations (obsid 19562, 20893) simultaneously and do not include the short, older observation as some of the model parameters (e.g. \textit{N$_{H,los}$}, the line of sight column density) can be time-dependent. 

\subsubsection{Spectral Modelling of EELRs}

Figure \ref{fig:arms_oiii_xray_smooth} shows the dashed white ellipses which indicate the regions used to extract the spectra of the North and South EELRs.
We fit combinations of APEC and CLOUDY models multiplied by the galactic absorption model (\textit{phabs}) to the extracted spectra in the soft (0.3-2 keV) band. The north and south regions were fit separately and we fit only the 2 longest observations as the emission from the EELRs are not well constrained in the short 10 ks observations (obsid 18080). We again bin the spectra to 5 counts/bin and use the wstat statistic due to the low numbers of total counts in each region.

\subsection{Chandra PSF Modelling}

We followed the standard Chandra science threads\footnote{\url{http://cxc.harvard.edu/ciao/threads/prepchart/}} to simulate the Point Spread Function (PSF) of each observation and energy band (0.5-1.5 keV, 1.5-3 keV, 3-6 keV and 6-7 keV) using the Chandra Ray Tracer (ChaRT) v2 \citep{carter2003chart}. For the source spectrum we used the \cite{Zhao2020} spectral model plus our best-fitting CLOUDY model of the soft X-ray emission below 2 keV. We simulate each observation individually as they have different off-axis angles and aspect solutions. We ran 1000 iterations of the simulation for each observation and energy band to ensure our PSF model was robust.

We used MARX \citep{2012SPIE.8443E..1AD} to project the simulated rays from ChaRT onto the detector and generate a PSF image. To test whether the apparent extent was real rather than an artefact of the PSF we extracted counts from both the observations and the simulated PSF images using Ciao's \textit{dmextract} tool. We use concentric annuli with radii 0$\farcs$3, 0$\farcs$6, 0$\farcs$9, 1$\farcs$2, 1$\farcs$5, 2$\arcsec$, 2$\farcs$5 and 3$\arcsec$ centered on the nuclear source location. It is well documented that the Chandra PSF exhibits an asymmetry at small scales, and we exclude this region of higher counts using the \textit{make\_psf\_asymmetry\_region} for each observation. The same background region as used in the spectral extractions is used here, and the uncertainties in the counts in each region are given by the Poission distribution with $\sigma_N = \sqrt{N}$, or by the Gehrels approximation in the low count (N$\leq$25) regime $\sigma_N = \sqrt{N+0.75}$.

We use this to compute radial surface brightness profiles which we normalise by the counts in the central bin. For observations and energy ranges which demonstrate statistically significant extended emission we repeat the process with annuli split into quadrants aligned in the cardinal directions. This allows investigation of potential azimuthal variation in the extended emission, which would be further evidence of an origin unrelated to Chandra PSF spread, which is radially symmetric to first order \citep{allen2004parameterization}.

\subsection{[O~{\sc iii}] to Soft X-ray Ratios}

There is a well known relationship between soft (0.5-2 keV) X-ray emission and [O~{\sc iii}] emission, which is often found to be coincident in extent and morphology \citep[see e.g.][]{2006A&A...448..499B,wang2011deep}. The ratio itself has been found to be inversely proportional to the ionization parameter of the gas \citep{2006A&A...448..499B}. To produce a spatially-resolved image of this ratio across NGC 5972 we convert our HST [O~{\sc iii}] image to energy flux units by multiplying by inverse sensitivity (\textit{PHOTFLAM} header keyword) and the effective width of the filter. The effective width of the FR505N tunable ramp filter (centered on 5156\AA) was calculated with the \textit{stsynphot} package to be 112.07\AA. We then smooth the image using a 5 pixel Gaussian blur to match the angular resolution of Chandra, and reproject it using \textit{reproject\_image} to match the binning of the soft X-ray image. 
To create an X-ray image in units of erg cm$^{-2}$ s$^{-1}$ we use the output image of \textit{merge\_obs} filtered to between 0.3 and 2 keV, with 1:1 pixel binning and multiply by the average energy of the events using \textit{dmimgcalc}. We then crop to the scale of the HST WFC-3, and use \textit{dmimgcalc} to calculate the ratio of the 2 images. We also repeat using a 3$\sigma$ adaptively smoothed Chandra image, made with \textit{dmimgadapt} which is designed to preserve small-scale features as well as flux. 

\subsection{Optical Spectra}
\subsubsection{Optical Emission Line Fitting}

The 2D HST spectra have both a spatial and spectral axis, which allows investigation of the line and continuum emission at positions across the [O~{\sc iii}] bubble. Figure \ref{fig:stis_loc} shows the orientation of the STIS extraction slit along the long axis of the bubble which allows us to probe the velocities and strength of line emission at different locations across the [O~{\sc iii}] bubble. 
The G750M filter, which contains the notable emission lines; H$\alpha$, [N~{\sc ii}]$\lambda 6585$\AA, [N~{\sc i}]$\lambda 6529$\AA \ and the [S~{\sc ii}]$\lambda6718/6733$\AA \ doublet, shows very little continuum emission above background. The G430L filter, which is dominated by the [O~{\sc iii}]$\lambda5008$\AA \ emission line, shows some evidence of weak continuum background. This background is low and constant across this emission line, so we fit it only as an additional additive constant during this analysis. We first shift the wavelength axis to account for the redshift of the galaxy due to the Hubble flow. The redshift of the galaxy inferred from the 21cm line of neutral hydrogen is z=0.02964 \citep{2014MNRAS.440..696A}. We then fit Gaussian profiles to these emission lines at different spatial offsets using the \textit{lmfit} package. The model finds the best fits and confidence intervals of the mean wavelength, Full-Width Half Maximum (FWHM) and amplitude for each line. The \textit{lmfit} package uses a Levenberg-Marquardt Least Squares minimisation algorithm \citep[see][]{marquardt1963algorithm} to find the best-fitting parameters and an F-test \citep{hahs2013introduction} to compute the confidence intervals.

The resultant fitted models are filtered to exclude models with extreme FWHM or very low amplitude. We also interactively exclude models that are a poor fit to the data. For each well-fitted emission line the Doppler shift of the observed mean wavelength from its rest vacuum wavelength\footnote{\url{http://classic.sdss.org/dr6/algorithms/linestable.html}} is used to calculate the velocity of the gas relative to the central offset with the strongest [O~{\sc iii}] flux. The velocity dispersion is also calculated, which is a measure of the velocity spread from the broadening of the emission lines. The effect of instrumental broadening is removed subtracting the velocity resolution of the filter in quadrature (16  km s$^{-1}$ for G750M, 98  km s$^{-1}$ for G430L\footnote{\url{www.stsci.edu/hst/instrumentation/stis/performance/spectral-resolution}}).

\subsubsection{Balmer Decrement Correction}
\label{sec:bpt}
In our data set both the [N~{\sc ii}] $\lambda$6585\AA \, H$\alpha$ and [O~{\sc iii}] $\lambda$5008\AA \ emission lines are very strong across the bubble, but the H$\beta$ line is much weaker. This means the background stellar continuum, which absorbs H$\beta$ flux, must be modelled and subtracted in order to accurately estimate the unabsorbed line flux.
To do this we use the \textit{Penalized Pixel-Fitting (pPXF)} package detailed in \cite{2017MNRAS.466..798C} which fits combinations of stellar templates to the spectra. We used the MILES library of stellar spectra \citep{2011A&A...532A..95F} based on the \cite{2010MNRAS.404.1639V} population models. We degraded the resolution of the Vazdekis templates (2.51\AA) to match the spectral resolution of the G43OL filter. The spectral lines are masked during the fit, but the fitted stellar populations are used to estimate the H$\beta$ flux absorbed by the stellar continuum. 

We subtract the continuum model from the spectra to separate out the emission lines, and then calculate the flux in the $H\beta$ line using \textit{lmfit}. We also determine the flux in the H$\alpha$ line, and calculate the central Balmer decrement (H$\alpha$/H$\beta$) assuming constant reddening across the bubble. We then used this relationship to estimate off-axis H$\beta$ fluxes. We then produce the common diagnostic diagram is known as a Baldwin, Phillips and Terlevich (BPT) diagram \citep{1981PASP...93....5B} used to determine whether the ionization source is a Seyfert II, LINER or an HII region. We calculate the [O~{\sc iii}] $\lambda$5008\AA / H$\beta$ and [N~{\sc ii}] $\lambda$6585\AA /H$\alpha$ ratios of emission line flux and plot them to determine the ionization source of different regions across the bubble. 

We also estimate the average electron density of the bubble from the [S~{\sc ii}]$\lambda$6718/6733\AA \ doublet using Equation \ref{eq:sii_density} \citep{2006agna.book.....O}. This density measure works in regions with density below 10$^6$ cm$^{-3}$. We assume an electron temperature of 10,000 K as the [O~{\sc iii}]$\lambda$4363\AA \ emission line is not strong enough to reliably estimate the electron temperature. 

\begin{equation}
    N_e = 10^2 \sqrt{T_e}\frac{R_{\textrm{[S~{\sc ii}]}}-1.49}{5.61-12.8R_{\textrm{[S~{\sc ii}]}}} \textrm{cm}^{-3}
    \label{eq:sii_density}
\end{equation} where $R_\textrm{[S~{\sc ii}]}$=F($\lambda$6718\AA)/F($\lambda$6733\AA), the ratio of the flux in the  emission lines of the [S~{\sc ii}] doublet \citep{2006agna.book.....O}. 

\section{Results}
\label{sec:results}

\subsection{X-ray Imaging}

\begin{figure}
	\includegraphics[width=\columnwidth]{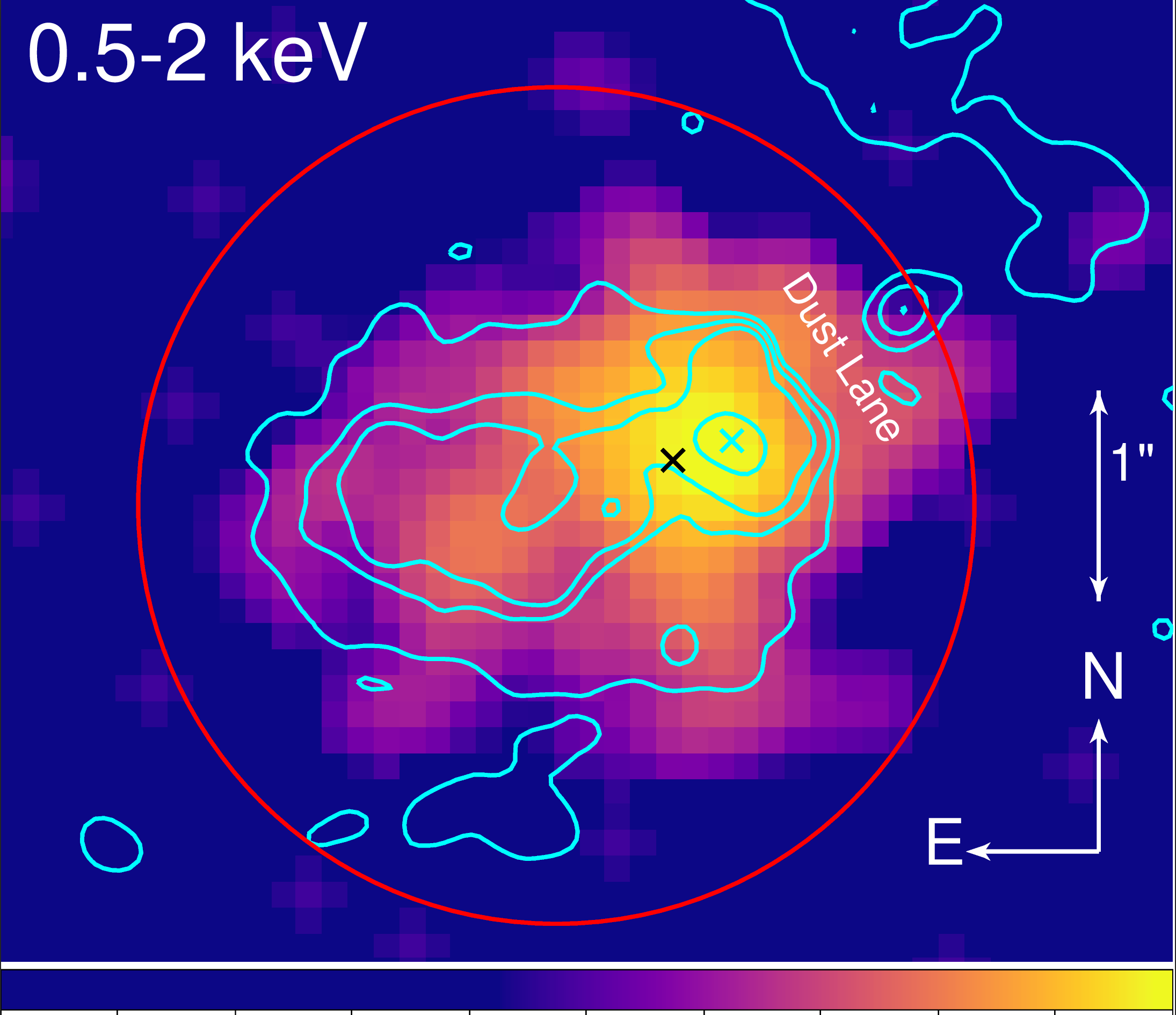}
    \caption{The smoothed soft (0.3 to 2 keV) X-ray observation binned to 1/4 ACIS-S pixel size of the central region of NGC 5972. The colour palette has been chosen to highlight the range of intensities across this region, and the X-ray data smoothed with a 3 pixel (3/4 ACIS pixel) Gaussian kernel. The spatially coincident HST [O~{\sc iii}] observation is shown with blue contours, with 5 contour levels and a dynamic range of 0.0015 to 0.01 photon s$^{-1}$ cm$^{-2}$ pixel$^{-1}$. The red circle is the same as shown in Figure \ref{fig:arms_oiii_xray_smooth} and the location of the SMBH is marked with a black cross. The labelled dust lane was identified by \protect\cite{Keel2015} from HST imaging.}
    \label{fig:bubble_xray_soft_smooth}
\end{figure}

\begin{figure*}
	\begin{subfigure}[b]{0.48\textwidth}
         \centering
        \includegraphics[width=\textwidth]{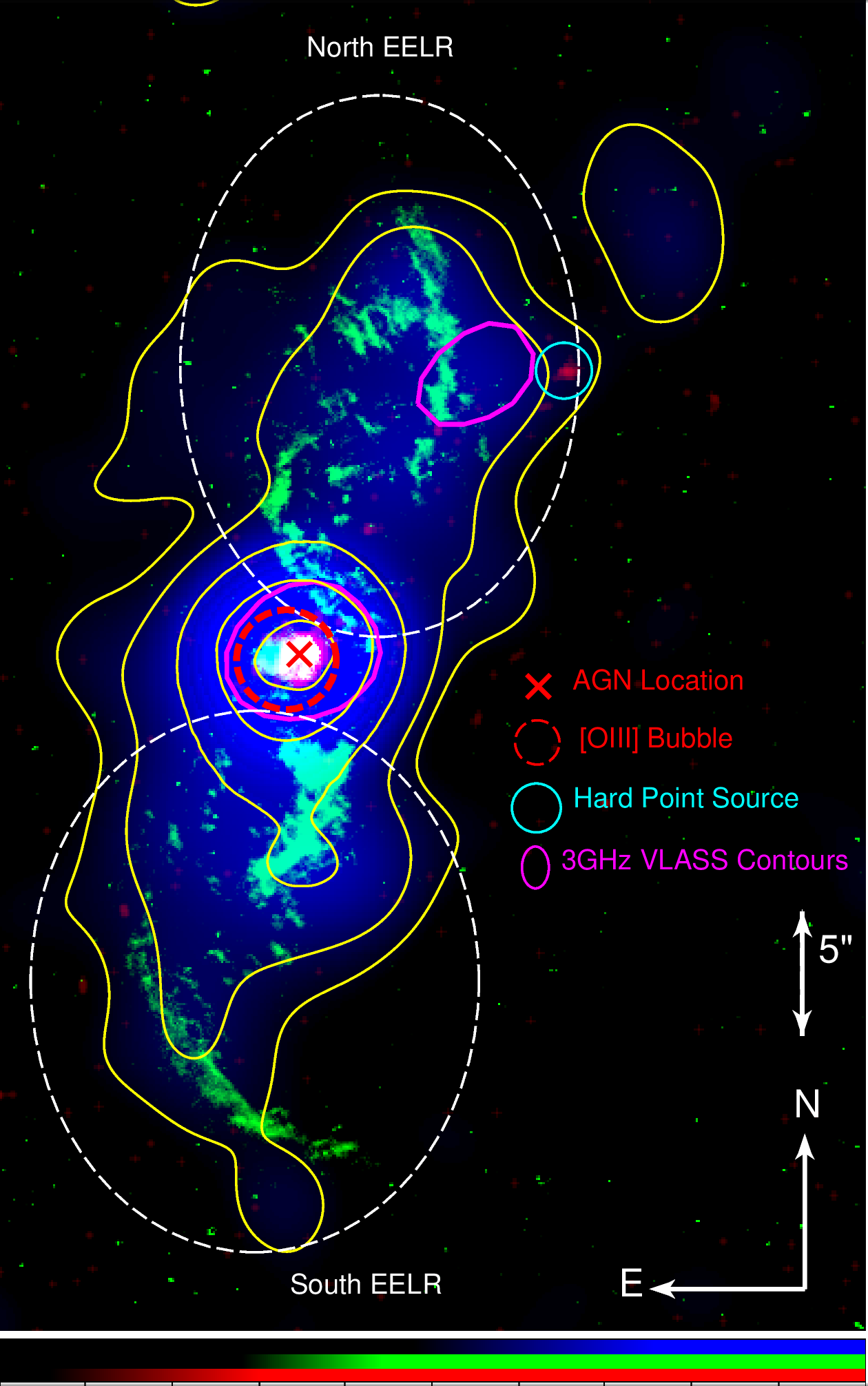}
     \end{subfigure}
     \hfill
     \begin{subfigure}[b]{0.48\textwidth}
         \centering
         \includegraphics[width=\textwidth]{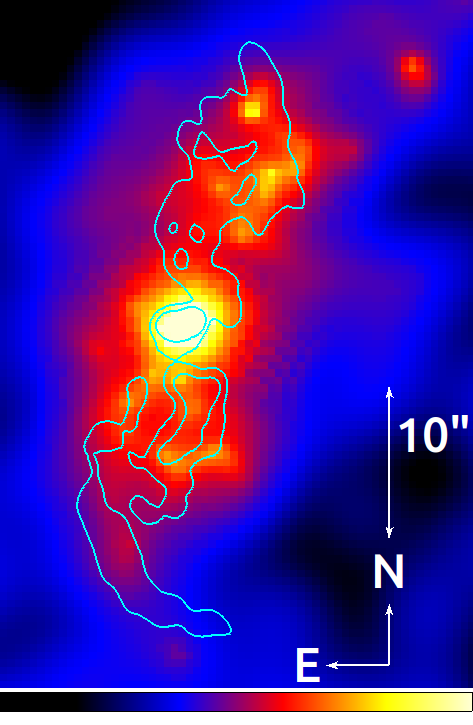}
     \end{subfigure}

    \caption{(\textit{left}) The HST [O~{\sc iii}] emission (green) overlaid with the soft (0.5-2 keV, blue), 3$\sigma$ adaptively smoothed ACIS-S X-ray observation of NGC 5972. The red channel represents the hard (2-8 keV) emission smoothed with a 3 pixel Gaussian. The logarithmically spaced contours (yellow) trace the soft emission, with 5 contour levels and a dynamic range of 0.01 to 0.5 counts cm$^{-2}$ s$^{-1}$ pixel$^{-1}$. The regions containing the EELRs are shown by dashed white ellipses, the [O~{\sc iii}] bubble location is circled in red and the nuclear source location is shown with a red cross. The purple contours show the location of the radio emission based on the 3 GHz VLASS observation. 
    (\textit{right}) The soft X-ray (0.5-2 keV) observation of NGC 5972, adaptively smoothed at 1$\sigma$, shown on a log scale. Cyan HST [O~{\sc iii}] contours of 0.004 and 0.019 counts cm$^{-2}$ s$^{-1}$ are overlaid to show the correlation with the EELR. }
    \label{fig:arms_oiii_xray_smooth}
\end{figure*}

\begin{figure}
	\includegraphics[width=\columnwidth]{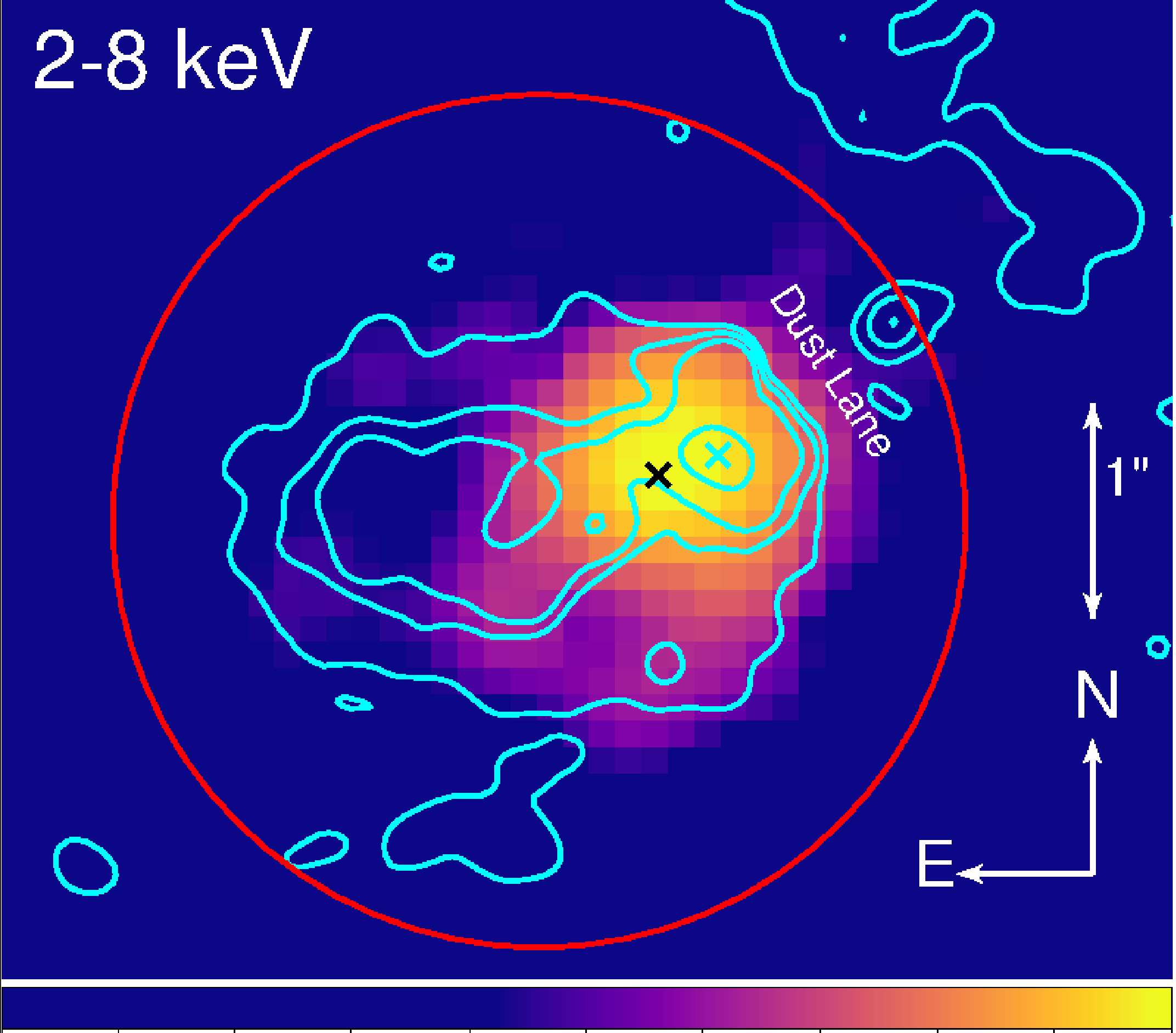}
    \caption{Hard 2-8 keV emission coincident with [O~{\sc iii}] bubble, shown with blue contours. Other Figure parameters are the same as Figure \ref{fig:bubble_xray_soft_smooth}.}
    \label{fig:bubble_xray_2-8keV}
\end{figure}

\begin{figure}
	\includegraphics[width=\columnwidth]{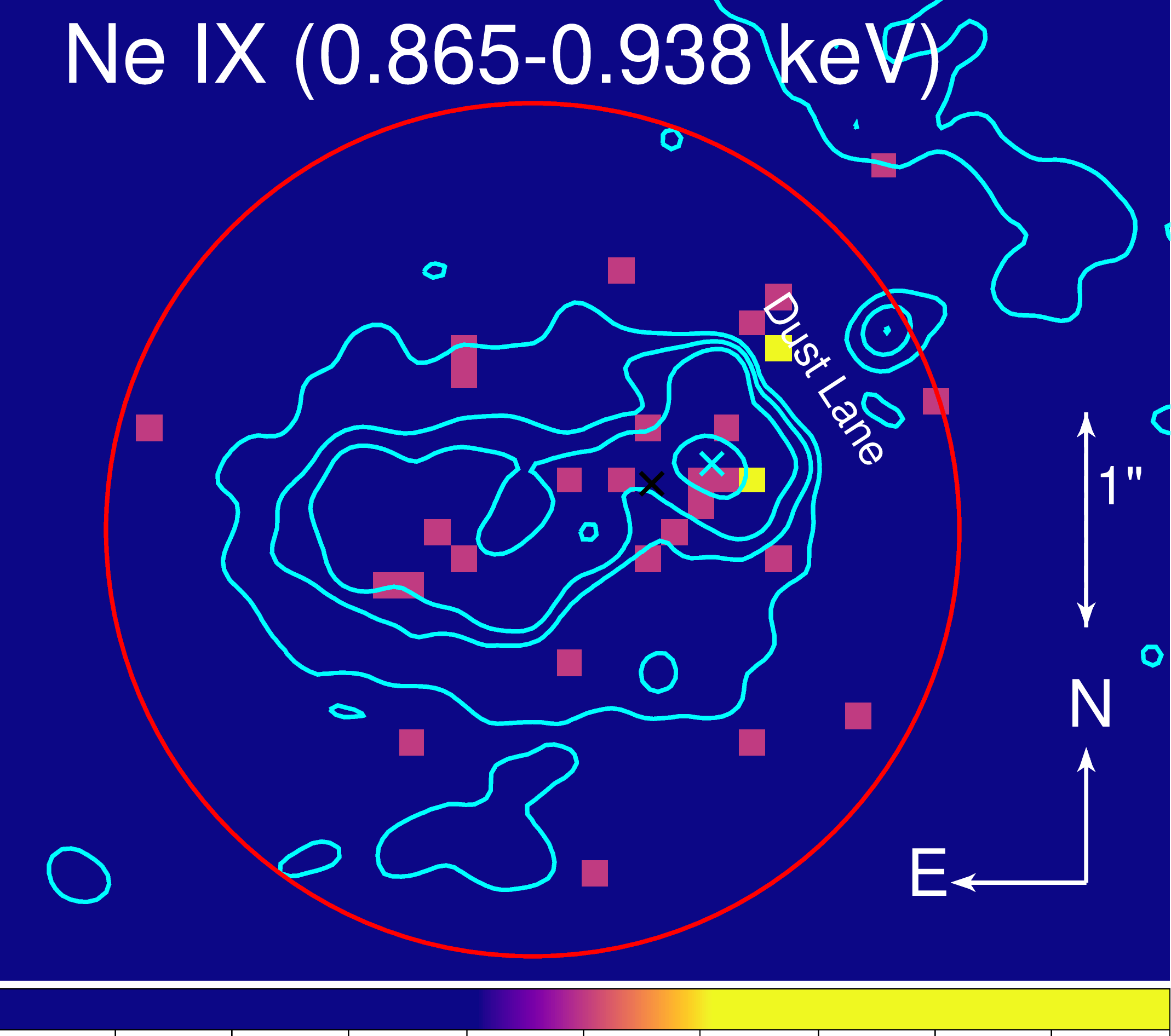}
    \caption{X-ray emission filtered to the 0.865-0.938 keV range containing the Ne IX emission line. Binned to 1/4 ACIS-s pixel size and shown with blue contours representing the coincident [O~{\sc iii}] bubble. X-Ray data is unsmoothed as there are a low number of counts in this narrow energy range. Other Figure parameters are the same as in Figure \ref{fig:bubble_xray_soft_smooth}. }
    \label{fig:bubble_xray_neix}
\end{figure}

\begin{figure}
	\includegraphics[width=\columnwidth]{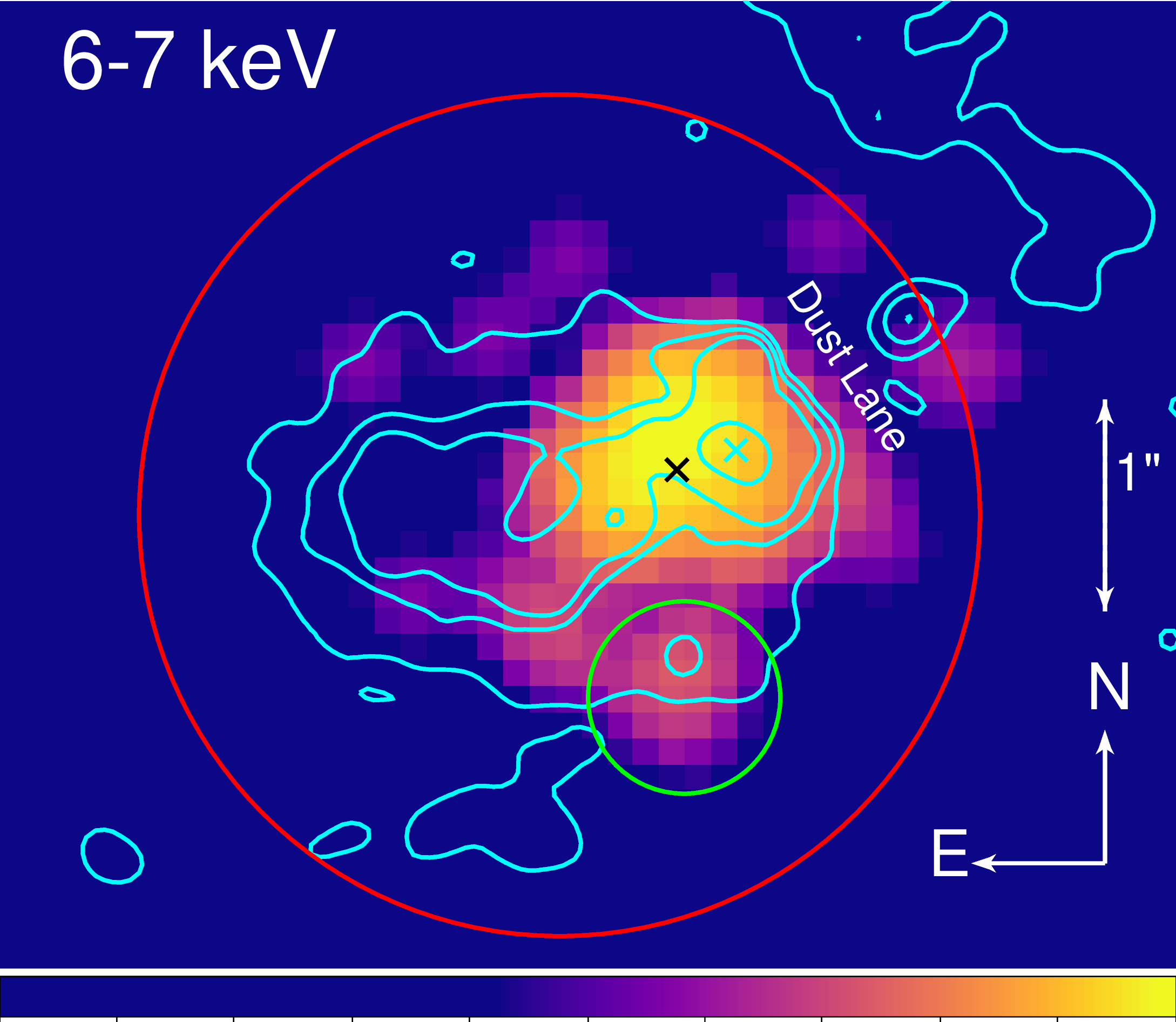}
    \caption{The smoothed hard (6 to 7 keV) X-ray observation binned to 1/4 ACIS-S pixels of the central region of NGC 5972. The colour palette has been chosen to highlight the range of intensities across this region, and the X-ray data smoothed with a 3 pixel (3/4 ACIS pixel) Gaussian kernel. The spatially coincident HST [O~{\sc iii}] observation is shown with blue contours. The red circle is the same as shown in Figure \ref{fig:arms_oiii_xray_smooth} and the location of the SMBH is marked with a black cross. The green circle shows the potential point source to the south of the nucleus. The labelled dust lane was identified by \protect\cite{Keel2015}. }
    \label{fig:bubble_xray_hard_smooth}
\end{figure}

\begin{figure}
	\includegraphics[width=\columnwidth]{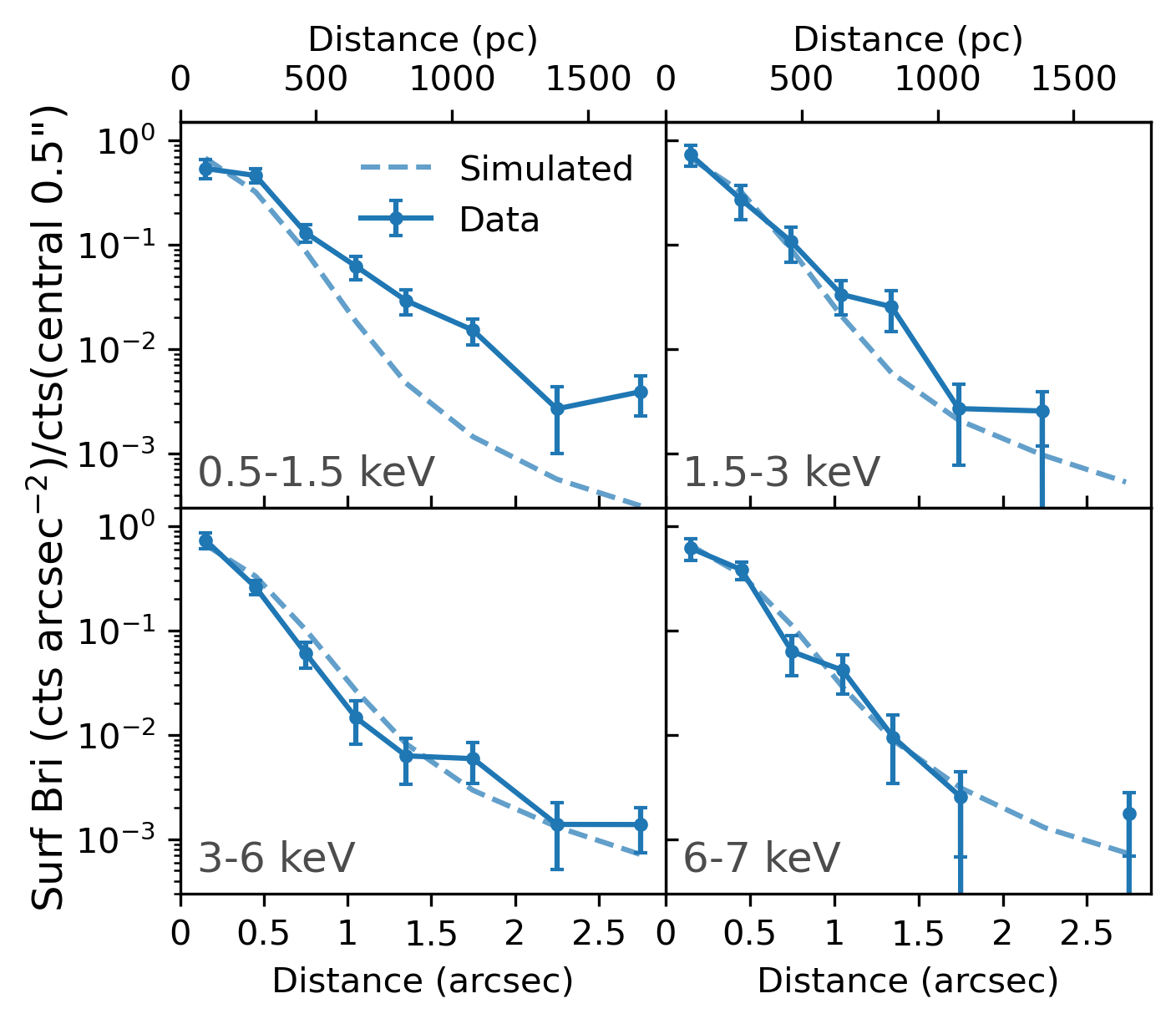}
    \caption{The radial surface brightness profile of the X-ray emission in different energy bands centered on the nuclear source location in Figure \ref{fig:arms_oiii_xray_smooth}. The profile for the real observations and simulations of the Chandra PSF are compared, and are normalised by the total number of counts within 0$\farcs$5 of the source. The counts are extracted from concentric circular annuli. The uncertainties in the PSF simulations are negligible in comparison to the observed data.}
    \label{fig:sur_brigh_all_energies}
\end{figure}

\begin{figure}
	\includegraphics[width=\columnwidth]{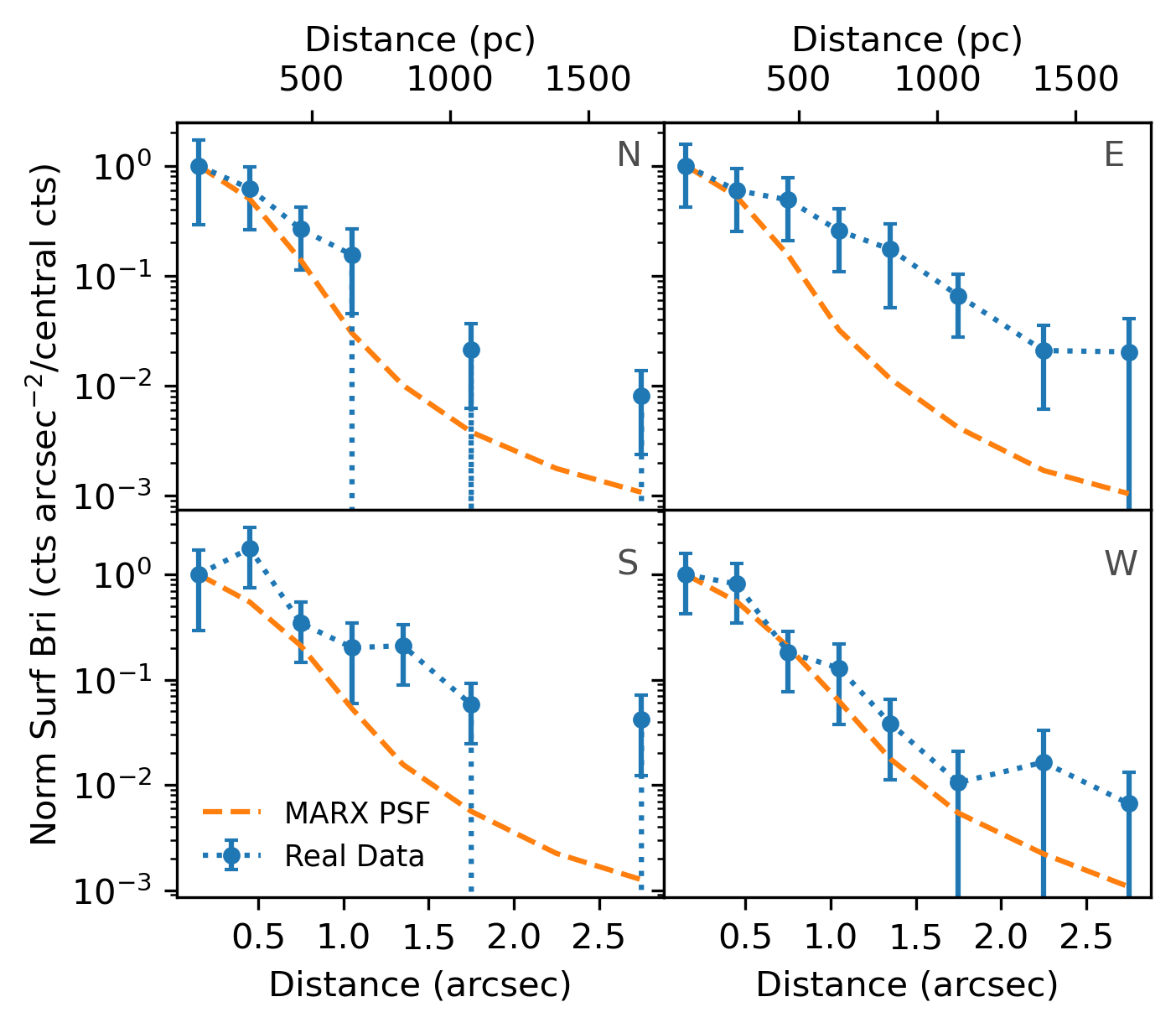}
    \caption{The surface brightness profiles of the soft (0.5-1.5 keV) X-ray emission split into quadrants aligned with the cardinal directions. The extracted brightness profiles centered on the nuclear source location (see Section \ref{sec:agn_loc}) and associated uncertainties are compared to the simulated PSF generated from ChaRT and MARX simulations. The surface brightness profiles are normalised by the counts within the central bin. The uncertainties in the PSF simulations are negligible in comparison to the observed data.}
    \label{fig:gradient_quadrants_soft}
\end{figure}

\subsubsection{Nucleus}
We report the detection of an above-background X-ray flux (0.3 - 8 keV) from the centre of NGC 5972. We measure a soft flux (0.5 - 2 keV) within a 2$\arcsec$ region of the central engine of of 2.65$\pm0.15 \times 10^{-14}$ erg cm$^{-2}$ s$^{-1}$ above background. This emission is spatially coincident with the bubble seen in the [O~{\sc iii}] HST observations. 

Figure \ref{fig:bubble_xray_soft_smooth} shows the smoothed soft (0.3 - 2 keV) X-ray emission spatially coincident with the [O~{\sc iii}] bubble. The false colour palette was chosen to highlight the surface brightness range across the emitting region. The HST [O~{\sc iii}] contours are overlaid for comparison, with the brightest [O~{\sc iii}] point labelled with a blue cross. The nuclear source location, shown with a black cross, is discussed in Section \ref{sec:agn_loc}. Figure \ref{fig:bubble_xray_2-8keV} shows the same comparison but for the hard broadband emission between 2 and 8 keV, smoothed with a 3 pixel Gaussian kernel. Figure \ref{fig:bubble_xray_hard_smooth} shows the smoothed 6-7 keV emission which is associated with the nuclear source. Counts within the asymmetry region have been removed as discussed in Section \ref{sec:agn_loc}. 
Figure \ref{fig:bubble_xray_2-8keV} isolates emission from the NeIX 0.905keV transition line (filtered to 0.865-0.938 keV), which has been associated with shocked emission in other sources (e.g. NGC 3933 \citealt{Maksym2019}, NGC 4151 \citealt{wang2011deep}). This image is shown unsmoothed with 1/4 ACIS pixel binning. 

\subsubsection{EELRs}
We also detect an above-background soft X-ray (0.3-2 keV) flux coincident with the EELRs to the north and south of the [O~{\sc iii}] bubble. This soft X-ray flux extends 15-20$\arcsec$ from the central region in both the North and South directions.
Figure \ref{fig:arms_oiii_xray_smooth} shows the white dashed ellipses we integrate the flux over for each EELR. We find a soft (0.5 - 2 keV) flux for the north (south) arm of 6.4$\pm$0.7 (5.4 $\pm$ 0.7) $\times 10^{-15}$ erg cm$^{-2}$ s$^{-1}$ above background. 
Figure \ref{fig:arms_oiii_xray_smooth} (left) shows the 3$\sigma$ adaptively smoothed soft X-ray (0.5-2 keV) in the blue channel overlaid on the green HST [O~{\sc iii}]. The red channel shows the hard (2-8 keV) emission, smoothed with a 3 pixel Gaussian. The soft X-ray shows the location of hot X-ray emitting gas relative to the EELR [O~{\sc iii}] emission and is further highlighted with yellow logarithmically-spaced contours. The location of the central engine and the region containing the [O~{\sc iii}] bubble are also shown with a red cross and dashed red circle. The nuclear source location is discussed in Section \ref{sec:agn_loc}. This image is smoothed to highlight the diffuse emission from the EELRs, and as a result the emission near the centre is purposefully over-exposed. Figure  \ref{fig:arms_oiii_xray_smooth} {right} shows the 1$\sigma$ adaptively smoothed soft X-ray (0.5-2 keV) emission overlaid with contours showing the  HST [O~{\sc iii}] EELR emission. 

We do not detect a significant hard X-ray flux coincident with either EELR, with the only exception being a potential hard X-ray point source in the North EELR. Figure \ref{fig:arms_oiii_xray_smooth} shows the potential point source with a blue circle. No optical counterpart is seen in HST narrowband or continuum imaging. This source is only marginally detected; we associate 6 counts between 3 and 6 keV with this source.

\subsubsection{Spatial Extent}

Figure \ref{fig:sur_brigh_all_energies} shows the radial surface brightness profile of the real and simulated observations for 4 different energy bands (0.5-1.5, 1.5-3, 3-6 and 6-7 keV) as a function of radial distance from the nuclear source location. The different observations are combined with an average weighted by exposure time, and the surface brightness is normalised by dividing each profile by the total number of counts within 0$\farcs$5 of the source location. The  $1\sigma$ uncertainties shown are propagated from the number of counts in each region under the Gehrels approximation. The Poisson uncertainty in the PSF simulations is negligible by comparison given the large number of simulations used. The soft emission (0.5-1.5 keV) clearly demonstrates spatial extent above the PSF up to 2$\farcs$5 from the central point. The 1.5-3 keV energy band also shows potential above background emission between 1$\arcsec$ and 1$\farcs$5 from the center. The 2 higher energy bands do not demonstrate any spatial extent when analysed radially. 
We also investigate the azimuthal dependence of the spatial extent of the soft emission. Figure \ref{fig:gradient_quadrants_soft} shows the surface brightness profile of the soft emission as a function of radial distance for quadrants aligned in the cardinal directions. The profiles are normalised by the central number of counts in each central quadrant.

\subsubsection{[O~{\sc iii}] to Soft X-ray Ratio Maps}

\begin{figure}
    \centering
    \includegraphics[width=\columnwidth]{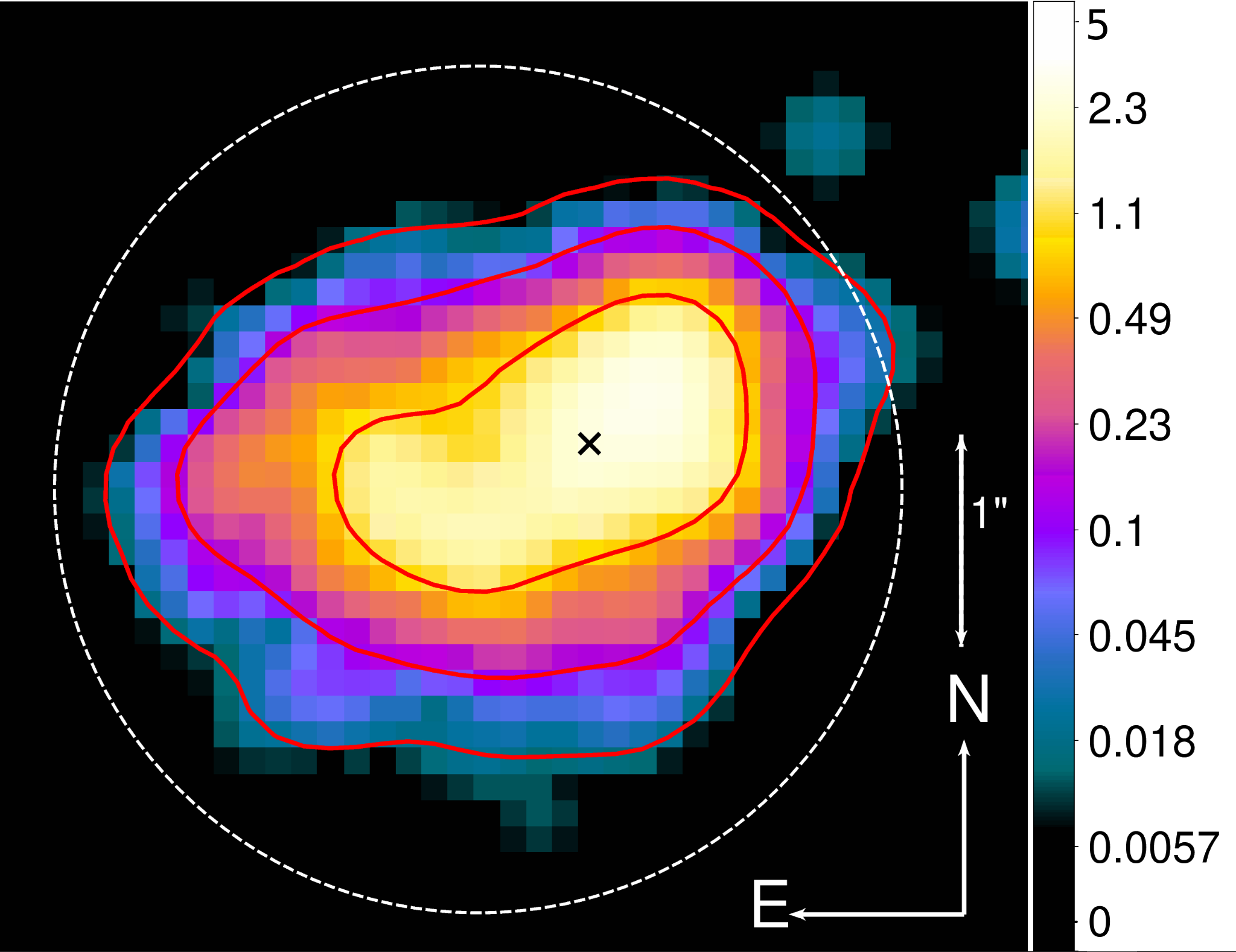}
    \caption{The ratio of F([O~{\sc iii}]$\lambda$5008\AA) to F(0.3-2keV) across the [O~{\sc iii}] bubble, binned to 1/4 native pixel size. The X-ray data has been smoothed with a 3 pixel Gaussian kernel. The black cross shows the location of the nuclear source. Contours show logarithmic variation in the ratio. }
    \label{fig:oiii_ratio_bubble}
\end{figure}

Figure \ref{fig:oiii_ratio_bubble} shows the ratio of [O~{\sc iii}]$\lambda$5008\AA \ flux to soft (0.5-2 keV) X-ray flux across the bubble, binned to 1/4 ACIS-S pixel size. The ratio map is smoothed with a 3 pixel Gaussian kernel. 

\begin{figure*}
    \centering
    \includegraphics[width=\textwidth]{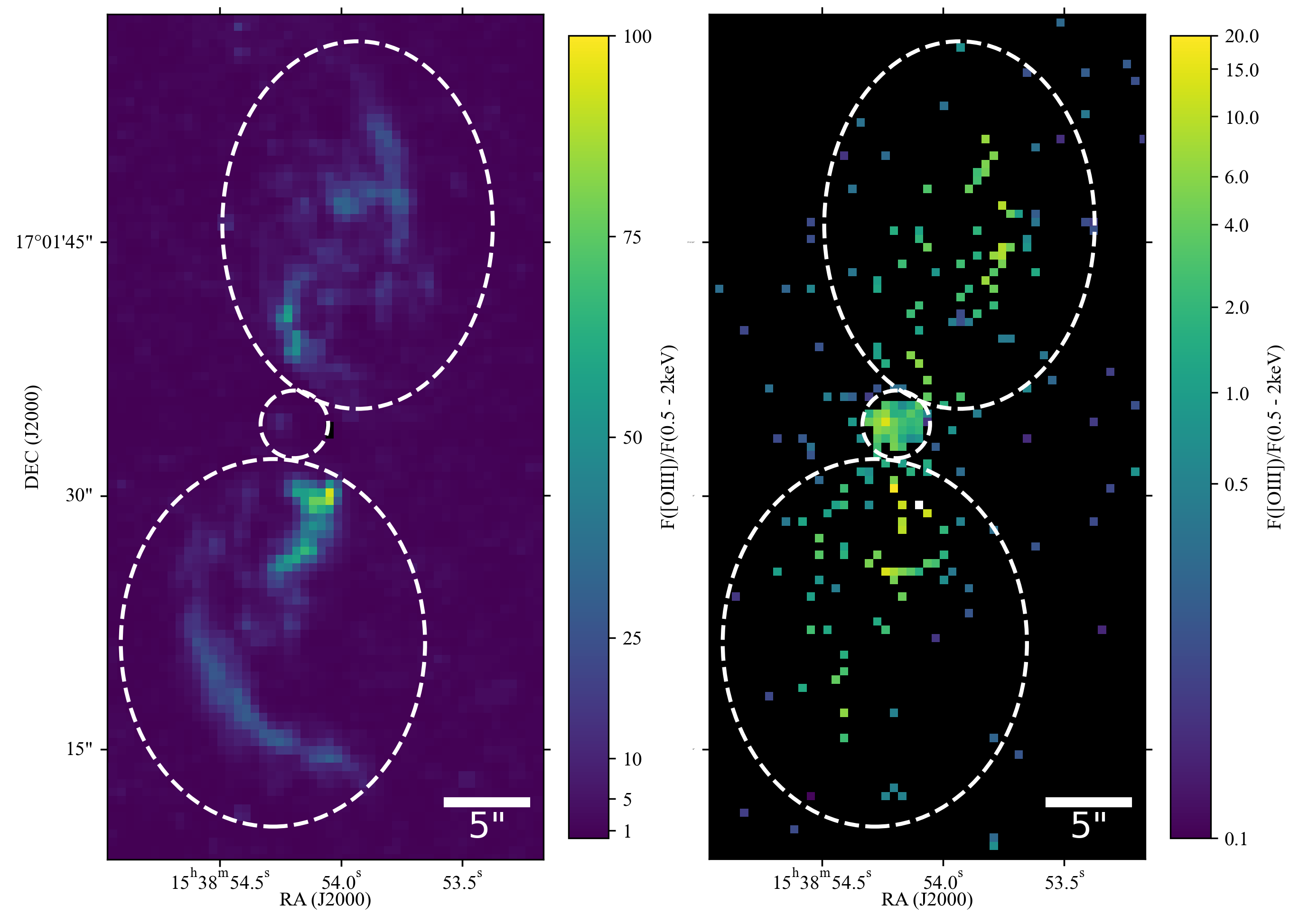}
    \caption{The ratios of F([O~{\sc iii}]$\lambda$5008\AA) to F(0.3-2keV) for the EELRs, binned to the native ACIS-S pixel size. In the left plot the soft X-ray band has been adaptively smoothed at 3$\sigma$ with \textit{dmimgadapt}, and in the right the raw unsmoothed ratio map is shown. Regions showing the location of the EELRs and [O~{\sc iii}] bubble are overlaid for comparison. In the unsmoothed image black pixels indicate no X-ray flux.}
    \label{fig:oiii_ratio_arms}
\end{figure*}

Figure \ref{fig:oiii_ratio_arms} shows the same ratio for the region covering both EELRs, which we bin to the ACIS-S pixel size given the low number of counts in the EELR regions. We show the ratio map using both 3$\sigma$ adaptively smoothed soft X-rays and the unsmoothed X-rays, which produce different flux ratio ranges. We indicate pixels with no soft X-ray flux in black.

From the regions we have defined we compute a [O~{\sc iii}]$\lambda$5008\AA \ flux from HST ACS image for the EELRs and [O~{\sc iii}] bubble. For the north (south) EELR we calculate 1.43$\pm$0.06 (1.72$\pm$0.07) $\times10^{-13}$ erg cm$^{-2}$ s$^{-1}$ above-background. For the [O~{\sc iii}] bubble we compute an [O~{\sc iii}] flux of 7.27$\pm$0.45 $\times10^{-13}$ erg cm$^{-2}$ s$^{-1}$.

Combining this with our computed X-ray fluxes we calculate average [O~{\sc iii}]/soft X-ray flux ratios which are less sensitive to single pixel variation. 
For the [O~{\sc iii}] bubble we compute a ratio of 2.67$\pm$0.23. For the north and south EELRs we compute ratios of  24.4$\pm$3.7 and 34.1$\pm$5.4 respectively.

\subsection{X-ray Spectroscopy}

\subsubsection{Nucleus}

\begin{figure*}
	
	\includegraphics[width=0.99\textwidth]{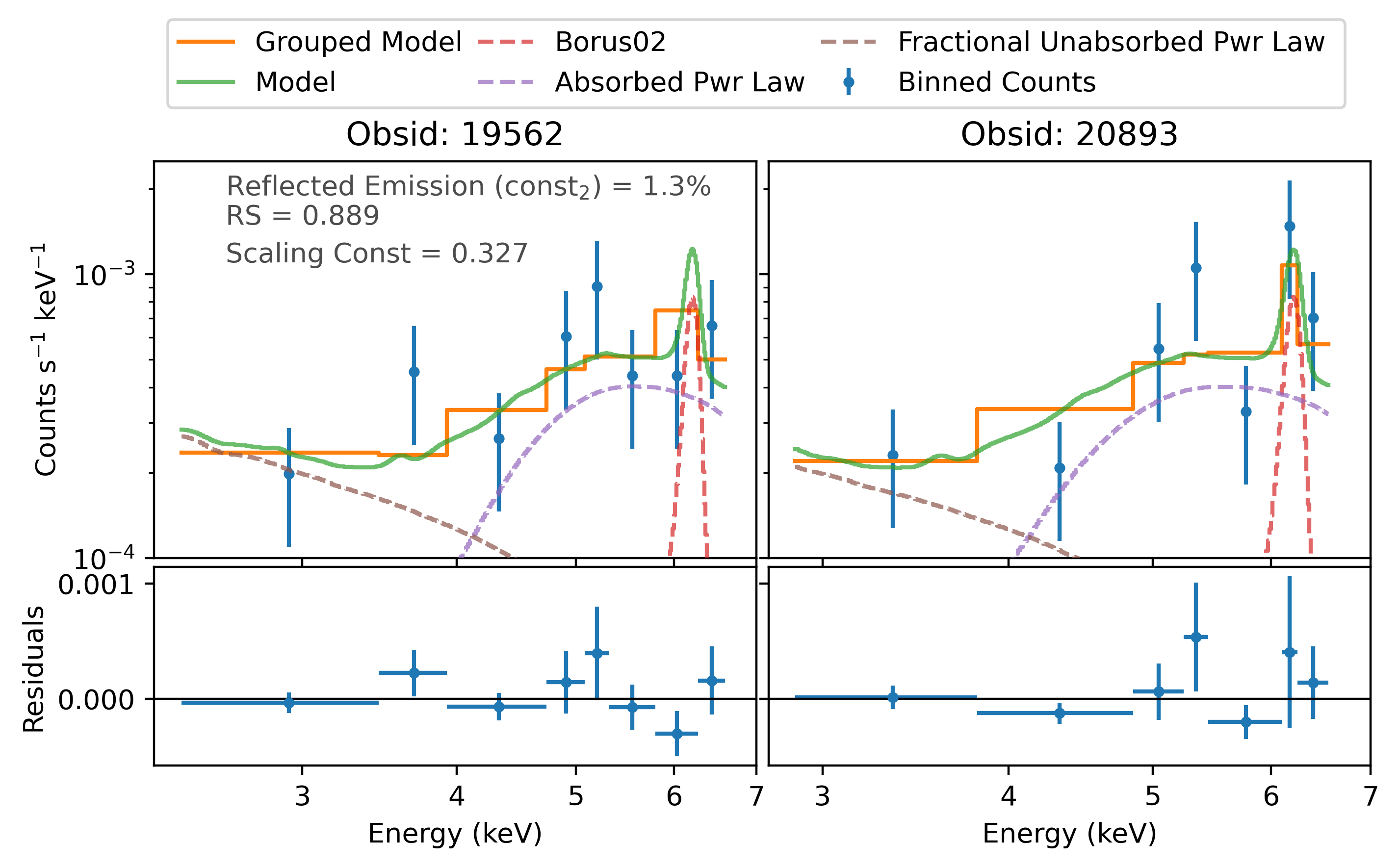}
	\includegraphics[width=0.99\textwidth]{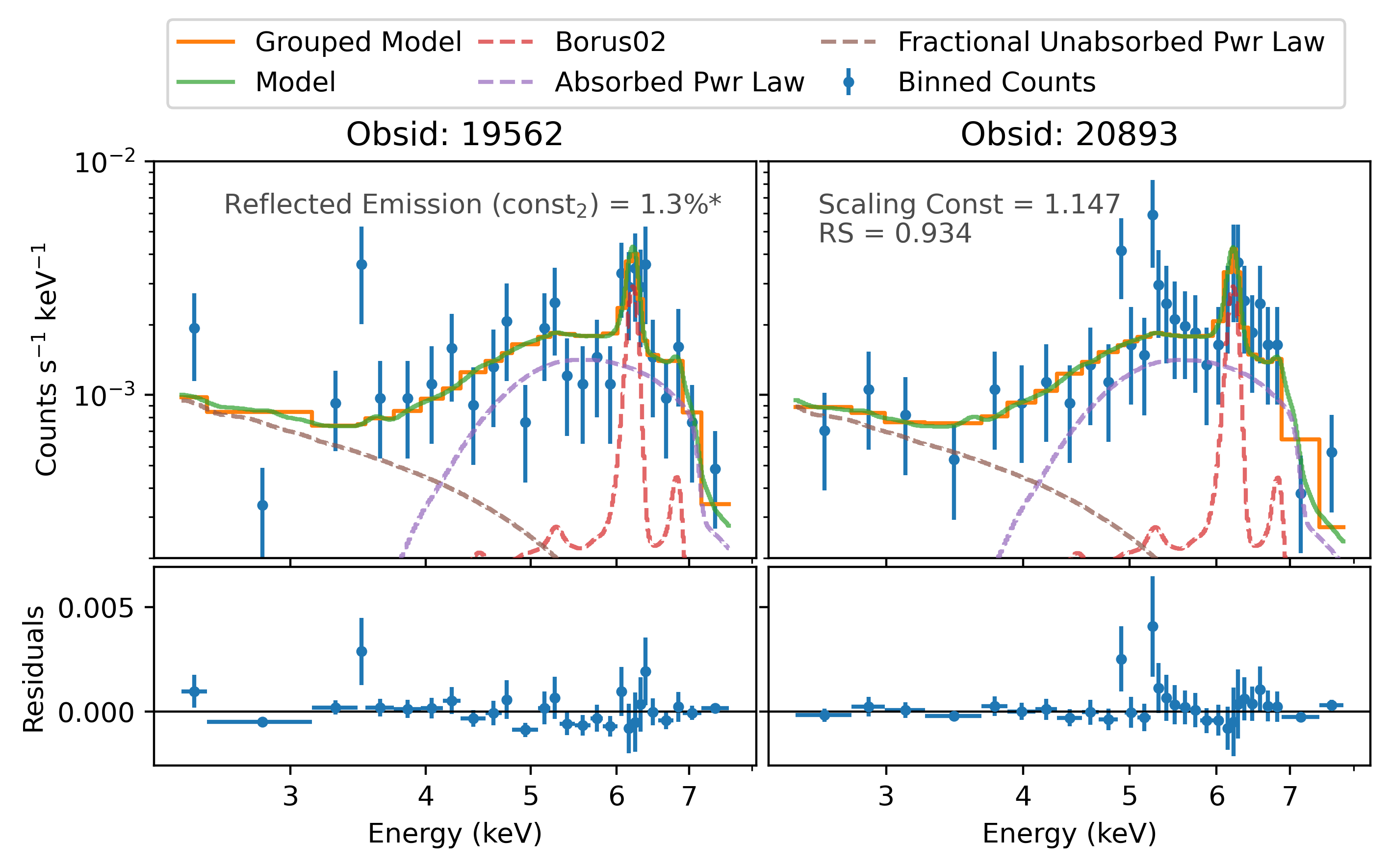}
    \caption{The hard (3-8 keV) spectra extracted from 0$\farcs$25 \textit{(top)} and 2$\arcsec$ \textit{(bottom)} regions centered on the nuclear source location, binned to 5 counts/bin, for 2 of the Chandra observations of NGC 5972. \protect\cite{Zhao2020}'s model is overlaid, with the model components shown in dashed lines. The residuals of the model for each observation are shown underneath. The values of the scaling constants are shown, with a * indicating if the component is frozen during the fitting process. }
    \label{fig:spectra_core}
\end{figure*}

The top plot in Figure \ref{fig:spectra_core} shows the hard spectra (2.8-8 keV) extracted from a 0$\farcs$25 region centered on the nuclear source location (see \ref{sec:agn_loc}). The spectra are fitted to the model described in Section \ref{sec:nucleus_spec} from \cite{Zhao2020}, with all parameters other than the 2 scaling constants fixed to the authors best-fitting values. We show the absolute residuals of this model for each observation under each spectra. From the fit we find a best-fitting value of 1.3$^{+0.6}_{-0.4}$\% for the scaling constant which represents reflected emission (const$_2$). The fit has a reduced statistic of 0.90, indicating a good fit to the data. For comparison \cite{Zhao2020} found a value of 1.4\%, which is within error of our value.

Const$_2$ was frozen at this value, and the bottom plot in Figure \ref{fig:spectra_core} shows the fit and residuals of this model to the spectra extracted from a larger 2$\arcsec$ region, which we use to constrain the value of the overall scaling constant of the nuclear spectral model. The energy range is still constrained to 2.5-8 keV to exclude the soft emission. For this scaling constant (const$_1$) we calculate a value of 1.15$^{+0.07}_{-0.06}$ with a reduced statistic of 0.93. 

\begin{figure*}
	 
	\includegraphics{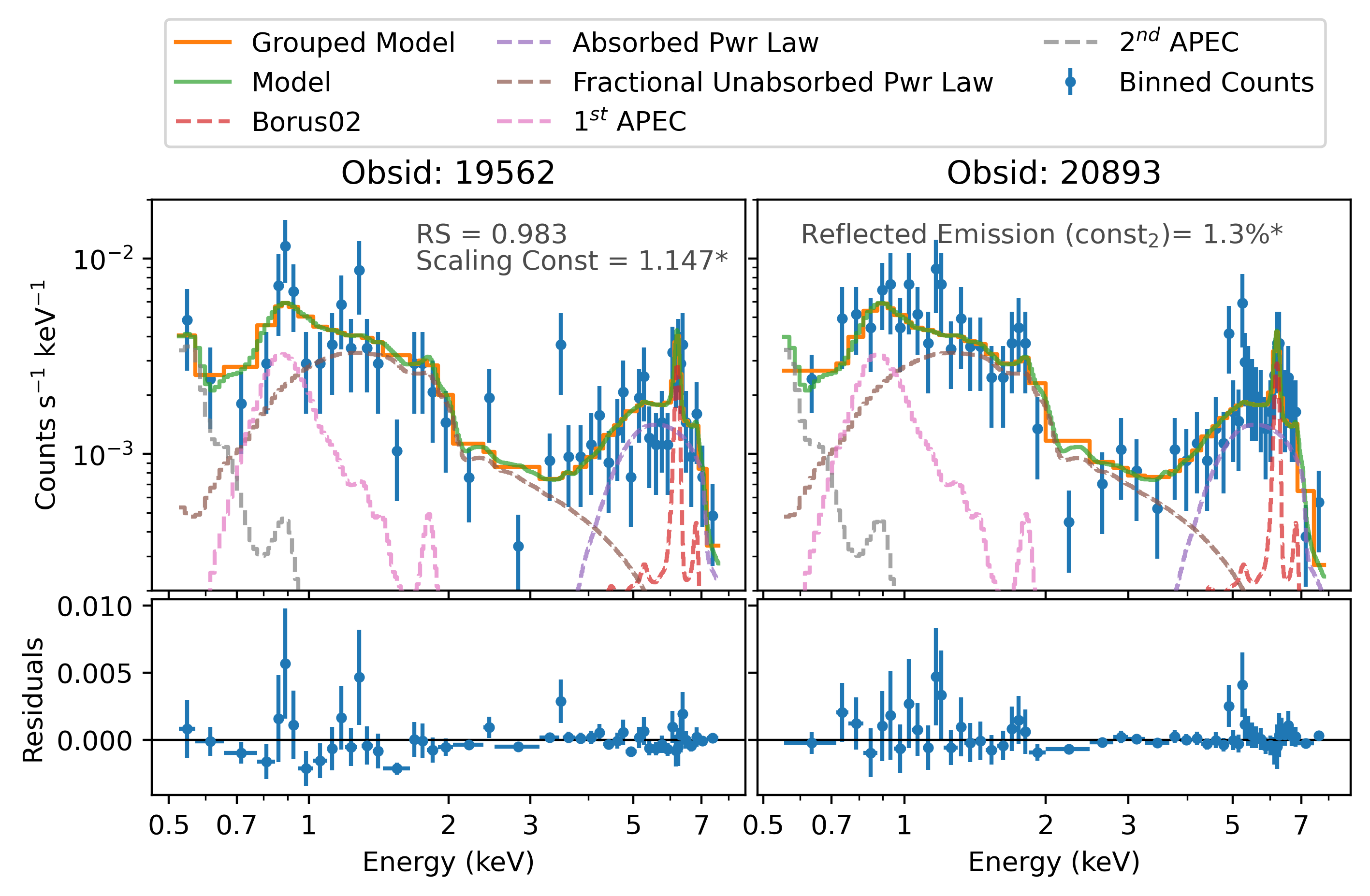}
    \caption{The 0.3-8 keV X-ray spectra extracted from a 2$\arcsec$ region centered on the nuclear source location, binned to 5 counts/bin, for 2 of the Chandra observations of NGC 5972. The best-fitting double APEC model is shown with grey and pink dashed lines. \protect\cite{Zhao2020}'s model is included for the harder emission. The residuals of the model for each observation are shown underneath. The goodness of fit statistic, overall scaling constant and value of constant$_2$ are shown. }
    \label{fig:spectra_bubble_2apec}
\end{figure*}

\begin{figure*}
	 
	\includegraphics[width=\textwidth]{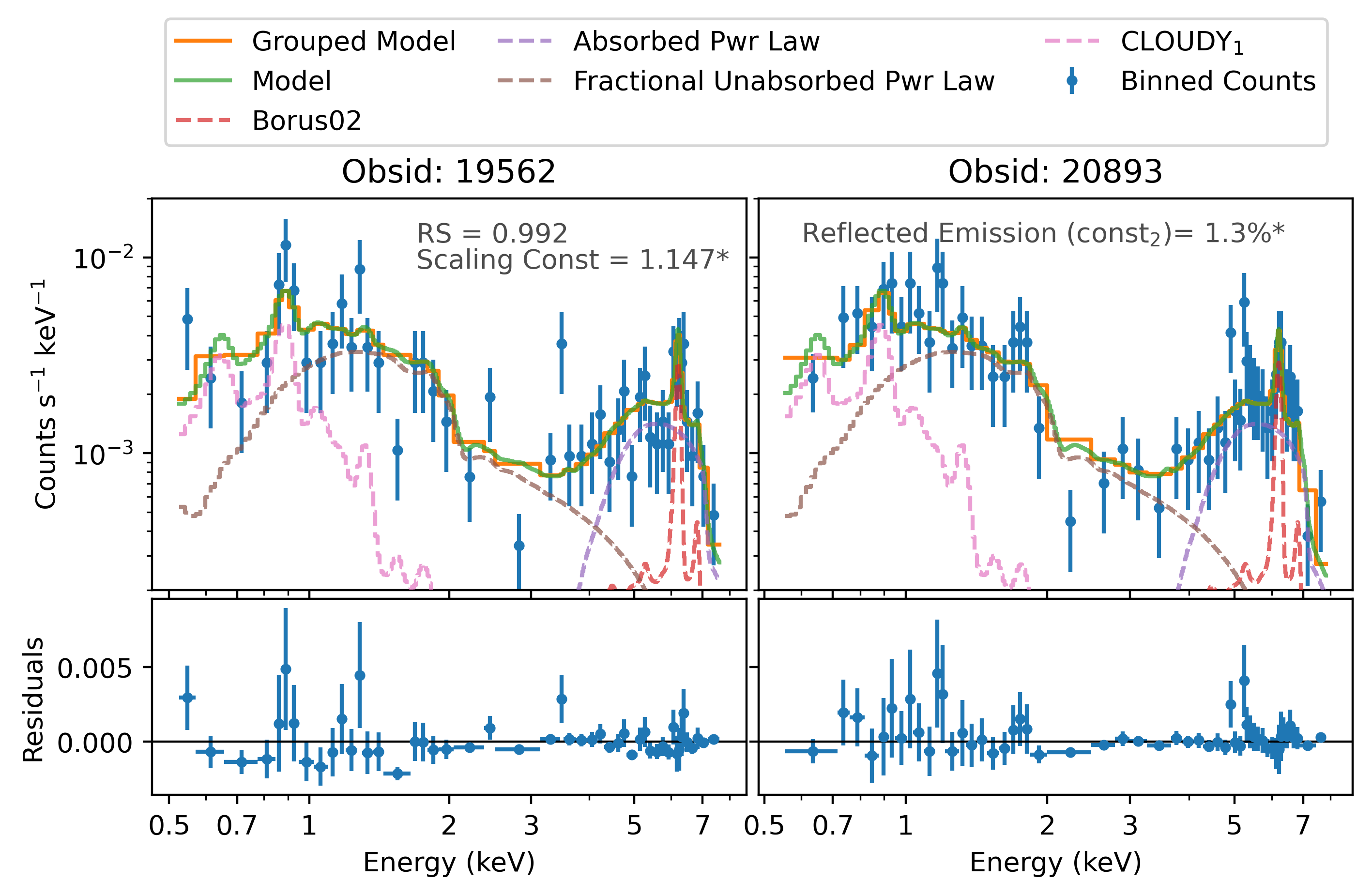}
    \caption{The 0.3-8 keV X-ray spectra extracted from a 2$\arcsec$ region centered on the nuclear source location, binned to 5 counts/bin, for 2 of the Chandra observations of NGC 5972. The pink dashed line shows the best-fitting CLOUDY model for the soft emission. \protect\cite{Zhao2020}'s model is included for the harder emission. The residuals of the model for each observation are shown underneath. The goodness of fit statistic, overall scaling constant and value of constant$_2$ are shown.}
    \label{fig:spectra_bubble_cloudy}
\end{figure*}

We then freeze both constants to produce a final model describing the nuclear emission above 2.5 keV. Using the same extraction region (2$\arcsec$ circle) we now expand our allowed energy range to include soft emission between 2.5 keV and 0.3 keV and introduce new models to fit these spectra. Table \ref{tab:soft_gas_models} shows the results of fitting combinations of APEC and CLOUDY models to this soft emission. The columns give the model used and the physically relevant model parameters - column density and ionization parameter for the CLOUDY model and temperature and emission measure (normalisation) for the APEC models. They also give the goodness of fit statistic, which is equivalent to the reduced Chi-Squared statistic, where a fit $\leq$1 indicates a good fit. 
Figure \ref{fig:spectra_bubble_2apec} shows the extracted spectra across the full energy range, fitted with our best double APEC model for the soft emission. 
Figure \ref{fig:spectra_bubble_cloudy} shows the 2 fitted spectra overlaid with the best CLOUDY fit. The goodness of fit statistic indicates a good fit but the best-fitting column density is at the lower  boundary for this model (log N$_{\textrm{H}}$ = 19.0 cm$^{-2}$). 

We have computed the bolometric luminosity of the AGN based on \cite{Zhao2020}'s estimate of the 2-10 keV luminosity from their X-ray spectral modelling. \cite{Zhao2020} found $6.17_{-1.15}^{+1.07} \times 10^{42}$ erg s$^{-1}$ for the 2-10 keV luminosity based on the same spectral model. Accounting for a bolometric correction factor of 12$^{+8}_{-3}$ taken from \cite{Netzer2019} we find a bolometric luminosity of $L_{\textrm{bol}}$ = $7.5_{-2.3}^{+5.2} \times 10^{43}$ erg s$^{-1}$.  The bolometric correction factor is not well-constrained, increasing the uncertainty in our luminosity estimate. Our bolometric luminosity estimate is consistent with the upper limit given by \cite{Keel2017} of <1.7 $\times 10^{44}$ erg s$^{-1}$ based on WISE Mid-IR observations. Finlez et al. 2022 (in preparation) find a current bolometric luminosity of $2.0 \times 10^{44}$ erg s$^{-1}$ based on Gemini IFU observations. 


\subsubsection{EELRs}
 
\begin{figure*}
	 \centering
	\includegraphics[width=\textwidth]{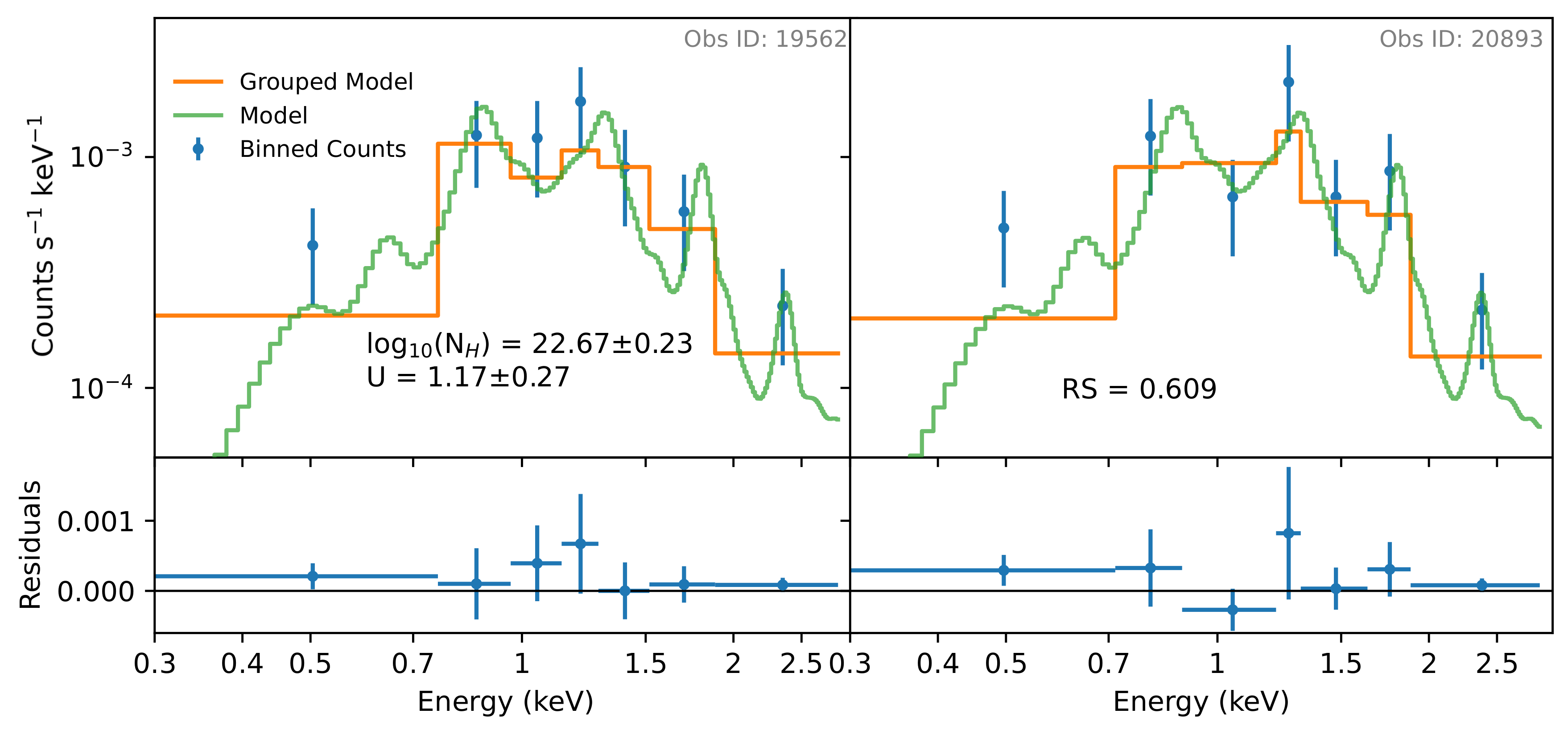}
    \caption{This figure shows the spectra extracted from the north EELR region shown by a dashed white ellipse in Figure \ref{fig:arms_oiii_xray_smooth} for 2 different observations. The spectra are binned to 5 counts/bin and the errorbars indicate the $1\sigma$ uncertainty under the Gehrels approximation. The best-fitting photo-ionization (CLOUDY) model is overlaid, both at its native resolution and binned to the same resolution as the spectra. The reduced statistic indicating the goodness of fit of the model, as well as the best-fitting parameters and their uncertainties are shown. The residuals of the model to each binned point are also shown in the lower plots. }
    \label{fig:north_arm_fit}
\end{figure*}

Table \ref{tab:arms_models} shows the results of fits of combinations of thermal (APEC) and photo-ionization (CLOUDY) models fitted to spectra extracted from the elliptical regions covering the north and south extended emission line regions. The best-fitting parameters and goodness of fit statistic are given. Figure \ref{fig:north_arm_fit} shows the best-fitting model and corresponding residuals for the spectra extracted from the north EELR. The best model is a single CLOUDY component multiplied by an absorbing \textit{phabs} model tied to the galactic absorption. This model is not a good fit to the spectra from the south EELR.

\begin{table*}
\renewcommand{\arraystretch}{1.5} 

\centering
\begin{tabular}{llll}
\textbf{Model}         & \textbf{Best Fit Parameters (Model 1)}     & \textbf{Best Fit Parameters (Model 2)}  & \textbf{Reduced Statistic~}   \\ 
\cline{1-4}
 CLOUDY                 & log N$_\textrm{H}$=19.0$^{+\textrm{u}}_{-\textrm{u}}$ cm$^{-2}$, U=0.32$^{+\textrm{u}}_{-\textrm{u}}$         &            & 0.99*                           \\
APEC                   & kT = 0.246$^{+0.04}_{-0.05}$ keV, norm = 2.4$^{+1.6}_{-0.7} \times 10^{-5}$                                  &                                                                                                        & 1.06~                           \\
 APEC + CLOUDY          & kT = 0.04 $^{+0.09}_{-0.02} $ keV, norm = 0.6 $^{+21}_{-0.6}$                    & log N$_{\textrm{H}}$= 19.4$^{+\textrm{u}}_{-\textrm{u}}$ cm$^{-2}$, U=0.48$^{+0.23}_{-0.30}$ & 0.99~                           \\
APEC$_1$ + APEC$_2$    & kT = $0.80^{+0.09}_{-0.12}$ keV ,~ norm =$4.8^{+1.8}_{-1.4}\times 10^{-6}$                                     & kT = $0.10^{0.06}_{-0.08}$ keV, norm = $4.2^{+\textrm{u}}_{-3.7} \ \times 10^{-4}$                              & 0.98~                           \\
CLOUDY$_1$+ CLOUDY$_2$ & log N$_{\textrm{H}}$=19.0$^{+\textrm{u}}_{-\textrm{u}}$ cm$^{-2}$, U=$0.32^{+\textrm{u}}_{-\textrm{u}}$       & log N$_{\textrm{H}}$=20.16$^{+\textrm{u}}_{-\textrm{u}}$ cm$^{-2}$, U=0.007$^{+\textrm{u}}_{-\textrm{u}}$   & 1.02*~                          \\

\end{tabular}

\caption{The best-fitting model parameters for combinations of spectral models and their corresponding goodness of fit statistic. The first column indicates the value of the constant which controls the strength of the reflected emission that dominates between 1 and 4 keV. A "*" next to the reduced statistic indicates that one or more model parameters is at a limit. "$\pm$u" indicates that the uncertainty on a parameter cannot be constrained.}
\label{tab:soft_gas_models}
\end{table*}

\begin{table*}
\renewcommand{\arraystretch}{1.5} 

\centering
\begin{tabular}{llll}
\textbf{Model}         & \textbf{Best Fit Parameters (Model 1)}      & \textbf{Best Fit Parameters (Model 2)}     & \textbf{Reduced Statistic~}  \\ 
\hline
North Arm              &                                                                                                        &                                                                                                         &                              \\ 
\hline
CLOUDY                 & log N$_\textrm{H}$=22.7$^{+0.2}_{-0.2}$ cm$^{-2}$, U=1.2$^{+0.3}_{-0.3}$                                &                                                                                                         & 0.61                         \\
APEC                   & kT = 3.4$^{+1.2}_{-1.2}$ keV, norm = 9.9$^{+2.0}_{-2.0} \times 10^-6$                                  &                                                                                                         & 1.38                         \\
APEC + CLOUDY          & kT = 0.012$^{+\textrm{u}}_{-\textrm{u}}$ keV ,~ norm =0.069$^{+0.06}_{-0.06}$                          & log N$_{\textrm{H}}$= 1.2$^{+0.3}_{-0.3}$ cm$^{-2}$, U=$22.8^{+0.4}_{-0.4}$                              & 0.74                         \\
APEC$_1$ + APEC$_2$    & kT = $3.4^{+\textrm{u}}_{-\textrm{u}}$ keV ,~ norm =$9.8^{+\textrm{u}}_{-\textrm{u}} \times 10^{-6}$   & kT = 0.008$^{+\textrm{u}}_{-\textrm{u}}$ keV ,~ norm =8$^{+\textrm{u}}_{-\textrm{u}} \times 10^{-4}$    & 1.65*                        \\
CLOUDY$_1$+ CLOUDY$_2$ & log N$_{\textrm{H}}$= 22.8$^{+\textrm{u}}_{-\textrm{u}}$ cm$^{-2}$, U=1.2$^{+\textrm{u}}_{-\textrm{u}}$ & log N$_{\textrm{H}}$= 19.6$^{+\textrm{u}}_{-\textrm{u}}$ cm$^{-2}$, U=0.07$^{+\textrm{u}}_{-\textrm{u}}$ & 0.83                         \\ 
\hline
South Arm              &                                                                                                        &                                                                                                         &                              \\ 
\hline
CLOUDY                 & log N$_{\textrm{H}}$=19 $^{+\textrm{u}}_{-\textrm{u}}$ cm$^{-2}$, U=$1^{+\textrm{u}}_{-\textrm{u}}$       &                                                                                                         & 1.43*                        \\
APEC                   & kT = 0.63$^{+0.12}_{-0.12}$ keV ,~ norm =$4.5^{+1}_{-1} \times 10^{-6}$                                &                                                                                                         & 1.3                         
\end{tabular}
\caption{Table contains best-fitting model parameters for X-ray spectra extracted from the north and south EELR regions in NGC 5972. The first column gives the spectra model fitted between 0.3 and 2.5 keV, with the second and third columns giving the best-fitting parameters of that model. All models were also multiplied by a fixed \textit{phabs} component representing galactic absorption. The last column gives the goodness of fit in the form of the reduced statistic calculated for each model. A star next to the reduced statistic indicates that one or more of the best-fitting parameters has reached a limit in the parameter space. Where possible uncertainty bounds are indicated on model parameters, but a $\pm$u indicates an parameter with unconstrained uncertainty.}
\label{tab:arms_models}
\end{table*}

\subsection{STIS Spectroscopy}
\subsubsection{Emission Line Profiles}
\begin{figure*}
    \centering
    \includegraphics[width=\textwidth]{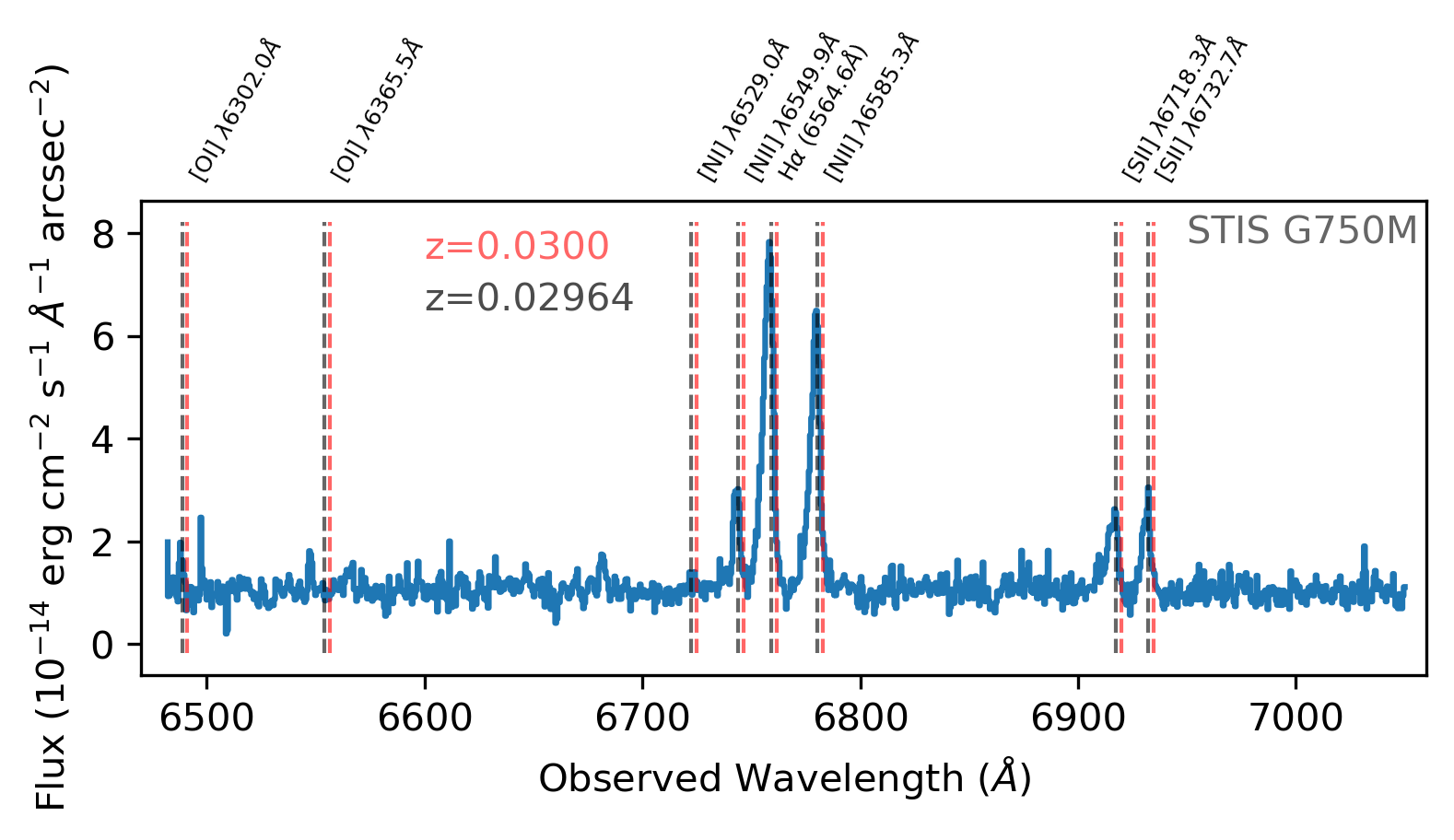}
    \includegraphics[width=\textwidth]{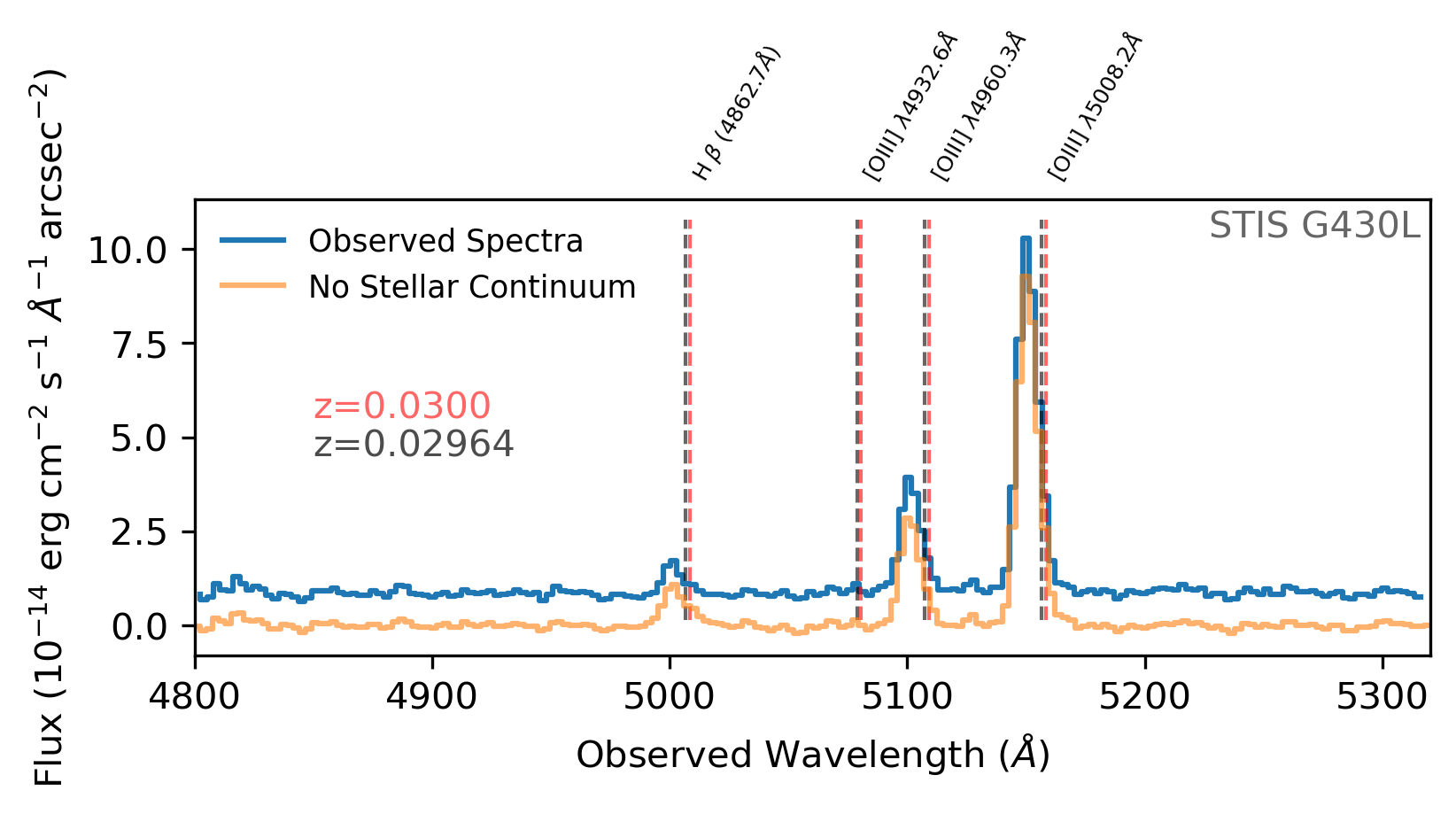}
    \caption{\textit{top (bottom)} STIS G750M (G430L) spectra of the nuclear bubble, summed across the 0$\farcs$5 region surrounding the central row of the 2D spectra. The observed wavelength is shown, as well as the locations of common emission lines at both the literature redshift and the Ca-triplet derived redshift. The emission lines are labelled with their names and rest wavelengths. For the G430L spectra both the observed spectra and the stellar continuum subtracted spectra from \textit{pPXF} are shown for comparison.}
    \label{fig:ap_spec}
\end{figure*}

Figure \ref{fig:ap_spec} shows the spectra extracted from the 2D STIS spectra along the central row, which is defined as the row with the strongest line emission. In pixel coordinates this corresponds to row y=626, 627 for the G750M and G430L observations respectively. The spectra are shown uncorrected for their redshift, but the theoretical wavelength of common emission lines are overlaid for both the literature redshift (z=0.0294) and the Ca-triplet derived redshift (z=0.0300) as discussed in Section \ref{sec:stis_kin_discuss}. The vacuum wavelengths are taken from the \textit{Sloan Digital Sky Survey} table\footnote{\url{http://classic.sdss.org/dr6/algorithms/linestable.html}}. The theoretical wavelengths show a systematic offset from the emission lines shown because of the relative velocity difference of the central region relative to the derived systemic velocity of the system. For the G430L spectra where we fitted the stellar continuum to find the unabsorbed H$\beta$ flux using \textit{pPXF}, the stellar continuum subtracted spectra is also shown.

\subsubsection{Kinematics}
\begin{figure}
	 
	\includegraphics[width=\columnwidth]{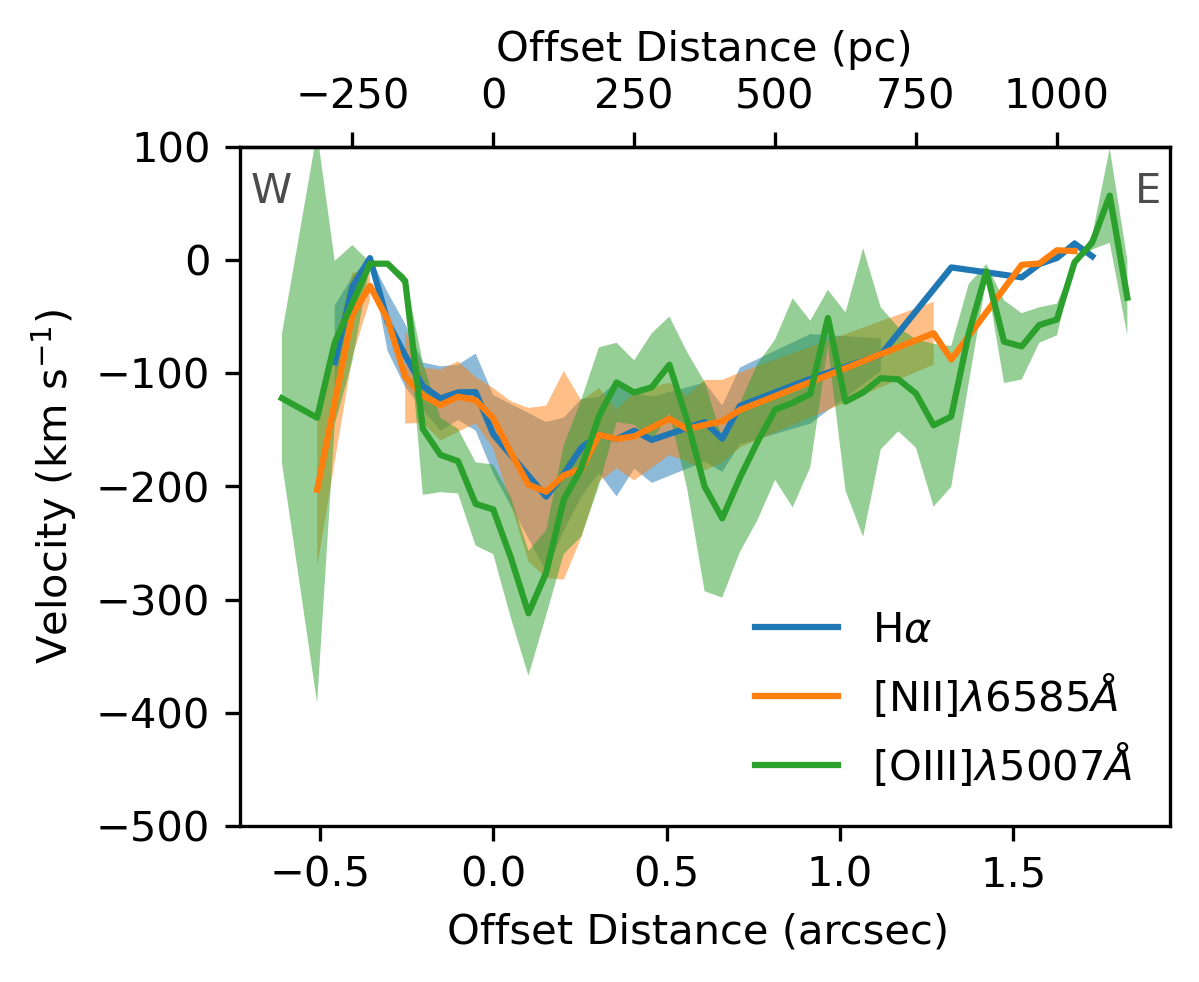}
    \caption{The radial velocity profile of the gas across the bubble computed from the Doppler shift of the bright [O~{\sc iii}], H$\alpha$ and [N~{\sc ii}] emission lines relative to their strongest points. The confidence intervals shown are propagated from the uncertainty in the Gaussian fitted to each emission line. The on the sky orientation of the extracted velocity profile is indicated by the cardinal directions in the upper right/left corners.}
    \label{fig:stis_velocity}
\end{figure}

Figure \ref{fig:stis_velocity} shows the computed radial velocity of the gas within the bubble as a function of offset position. The radial velocity was computed from the Doppler shift of the H$\alpha$, [N~{\sc ii}] and [O~{\sc iii}] emission lines relative to their rest wavelengths. The offset position (x-axis) was determined relative to the spectrum with the highest [O~{\sc iii}] line flux, which we have assumed to correspond to the brightest [O~{\sc iii}] region in the HST imaging.

\begin{figure}
	 
	\includegraphics[width=\columnwidth]{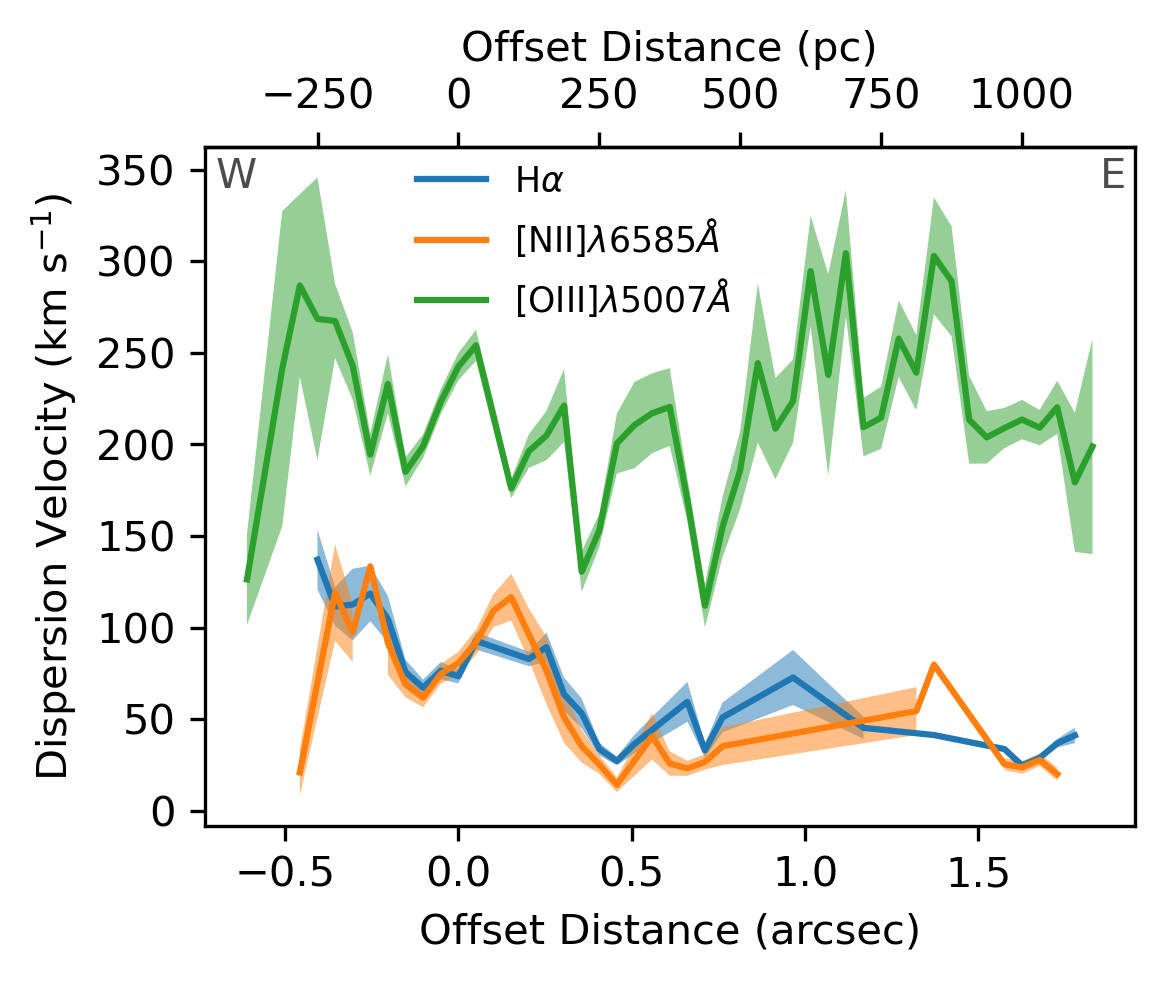}
    \caption{The velocity dispersion of the gas across the bubble computed from the Doppler broadening of the bright, H$\alpha$ and [N~{\sc ii}] emission lines relative to their strongest points. The instrumental broadening of each filter has been subtracted in quadrature. The confidence intervals shown are propagated from the uncertainty in the Gaussian fitted to each emission line. The on the sky orientation of the extracted velocity profile is indicated by the cardinal directions in the upper right/left corners.}
    \label{fig:stis_disp_velocity}
\end{figure}

Figure \ref{fig:stis_disp_velocity} shows the computed velocity dispersion of the gas as a function of offset position. Doppler broadening was used to compute the velocity dispersion from the FWHM of H$\alpha$ and [N~{\sc ii}] emission lines. We don't show the [O~{\sc iii}] velocity profile because it is dominated by the 300  km s$^{-1}$ instrumental broadening of the G430L filter. The confidence intervals are propagated from the uncertainty in the best-fitting of the FWHM of each emission line. 
\subsubsection{Density}
As discussed in Section \ref{sec:bpt} we calculate an average density of the [O~{\sc iii}] bubble using Equation \ref{eq:sii_density}, assuming an average electron temperature of 10,000 K. The doublet is not well-resolved across the bubble, so we average across the spatial axis to increase the signal. We calculate a ratio of 1.08$\pm$0.11. We then calculate an electron density of 500$\pm$220 cm$^{-3}$, averaged across the slit. The relationship is insensitive at this ratio, which causes large uncertainty in the density. 

\subsubsection{BPT Mapping}

\begin{figure}
	\includegraphics[width=\columnwidth]{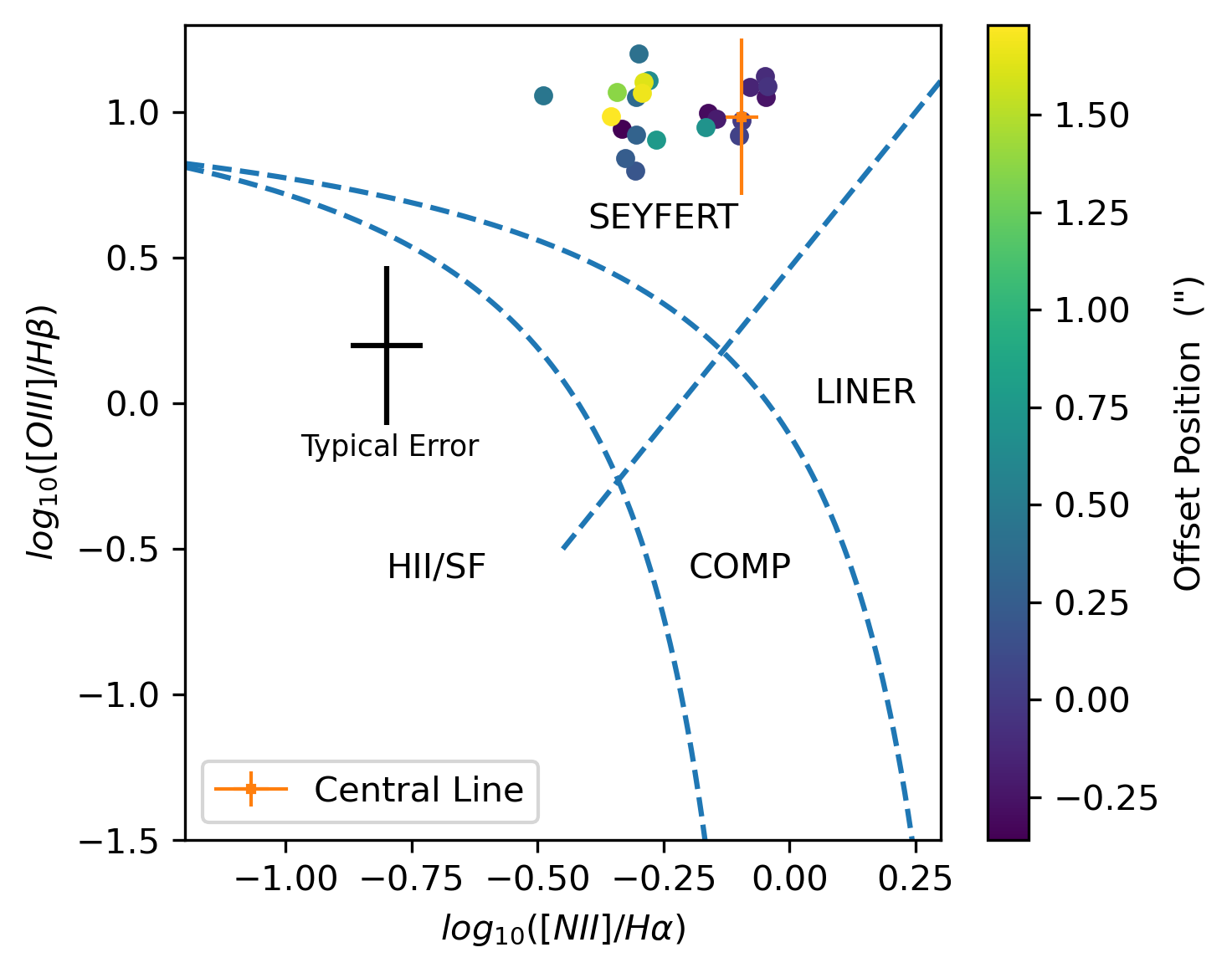}
    \caption{Baldwin, Phillips and Terlevich (BPT) diagram \citep{1981PASP...93....5B} showing the ratios of diagnostic lines as a function of offset position across the [O~{\sc iii}] bubble. The ratio of these lines indicates the ionising source, which are labelled in black and separated by the shaded blue lines. See section \ref{sec:bpt} for more details. The offset position is indicated by the colorbar on the right side of the plot. Due to the weakness of the H$\beta$ line, the Balmer decrement calculated at the central offset position is used as a proxy for H$\beta$ across the bubble, assuming constant reddening across the bubble. For comparison the H$\beta$ line strength at the central position is shown in orange. The dashed boundaries separating the different ionising sources are taken from \protect\cite{1981PASP...93....5B} and \protect\cite{2003MNRAS.346.1055K}.}
    \label{fig:bpt}
\end{figure}

From our fitting of the optical continuum using pPXF we find a best fit with the contributions from 3 stellar populations of distinct ages; a dominant component of old stars (age approximately 15 Gyr) and two young stellar populations of ages 0.4 and 0.06 Gyr. Subtracting this model for the stellar continuum and fitting the central H$\alpha$ and H$\beta$ lines we find a Balmer decrement of 3.6$\pm$2.2, which we use as proxy for H$\beta$ flux.
Figure \ref{fig:bpt} shows the Baldwin, Phillips and Terlevich (BPT) diagnostic diagram \citep{1981PASP...93....5B} for different offset regions across the [O~{\sc iii}] bubble. The offset of a particular point from the central position (highest flux [O~{\sc iii}] offset) is shown by the marker colour. As the H$\beta$ flux for offsets across the bubble is inferred from the H$\alpha$/H$\beta$ Balmer decrement at the central offset, this point is also plotted in orange for comparison. The average uncertainty for both ratios is shown. There is no clear trend across the [O~{\sc iii}] bubble in either of the ratios. All the emission from regions within the [O~{\sc iii}] bubble is consistent with a emission from a AGN (Seyfert) source. These results agree with Finlez et al. 2022 (in preparation) who find a stellar population within the bubble with an age of approximately 12 Gyr. The authors also find that the bubble and EELRs are consistent with ionization from AGN continuum emission.

\subsection{Keck Imaging}
\begin{figure}
	\includegraphics[width=\columnwidth]{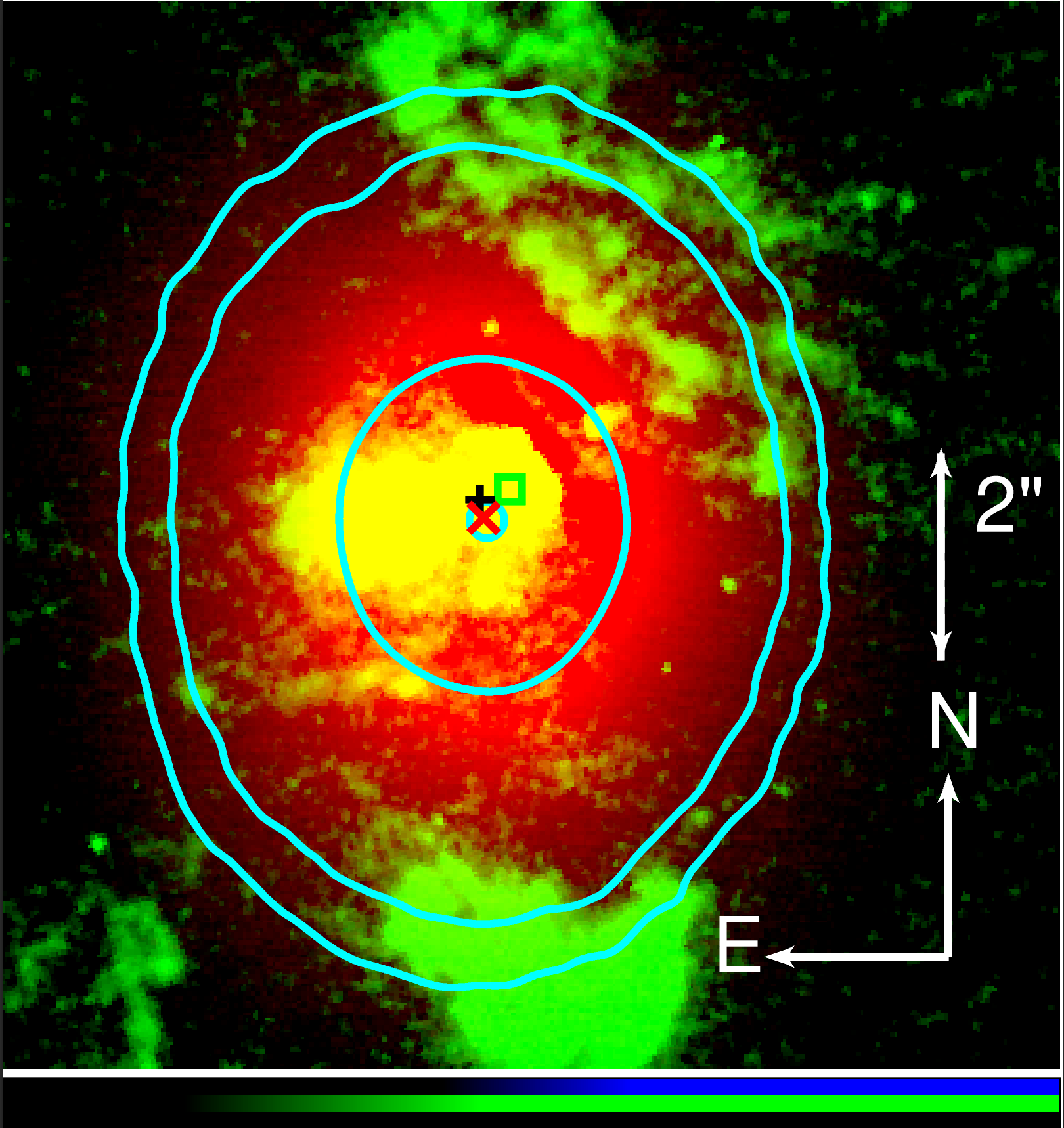}
    \caption{The Near-IR emission of NGC 5972 in the Kp band (shown in red) taken with the NIRC2 imager on the Keck II Observatory, overlaid on the HST [O~{\sc iii}] emission (shown in green). The Near-IR emission is shown with 4 contour levels and a dynamic range of 200 to 5000 counts. The nuclear source location is shown with a black plus, the centroid of the near-IR emission is shown by the red cross and the brightest [O~{\sc iii}] emission is shown with a green box. }
    \label{fig:bulge}
\end{figure}
Figure \ref{fig:bulge} shows an image of the near-IR emission of NGC 5972 taken with the NIRC2 imager on the Keck II Observatory. The near-IR emission is shown overlaid on the HST [O~{\sc iii}] emission.
The PSF achieved is 0$\farcs$15, based on nearby stars within the image. The nuclear source location is shown with a black plus, the centroid of the near-IR emission is shown by the red cross and the brightest [O~{\sc iii}] emission is shown with a green box. Given that the K band has an order of magnitude lower extinction when compared to the optical and also traces older populations from the stellar bulge, these NIR AO observations exclude the possibility of a larger separation (e.g. >100 pc) obscured nucleus.

\section{Discussion}
\label{sec:discussion}
\subsection{Nuclear Source Location and [O~{\sc iii}] Mismatch}
\label{sec:agn_loc}
The 6-7 keV emission from the neutral (Fe K$\alpha$) and ionized (Fe~{\sc xxv}, Fe~{\sc xxvi}) iron lines are thought to be emitted primarily by the outer accretion disk and inner region of the obscuring torus of the SMBH \citep[see e.g.][]{reynolds2012constraints,gandhi2015dust}. We expect this parsec scale emission to be point-like given the angular scale $\sim$600pc arcsec$^{-1}$ and the 0$\farcs$1-0$\farcs$2 resolution of these observations. The centroid of this emission is generally used as a signature of the central engine \citep[e.g.][]{maksym2017cheers}. The exact location of the central engine has consequences for much of the other analysis, including the spectroscopic and surface brightness profiles.

Specifically, as can be see in Figure \ref{fig:bubble_xray_hard_smooth}, there is a grouping of $\sim$9 counts $\sim$1$\arcsec$ to the south of the central region which appears to be an almost unresolved point source. This appears in multiple event files, so it is not an artefact of stacking misaligned observations. Further analysis suggests this is
(largely) an artefact of PSF asymmetry, which we quantify as follows. To determine the true location of the central engine we devised the following approach, which we apply separately to each observation.

We produced an image binned to 1/8 native pixel size (0$\farcs$06) filtered to 6-7 keV. We calculate the 50\% encircled energy using the \textit{mkpsfmap} tool, which is equivalent to 6 pixels ($\approx$0$\farcs$35) at this binning. We then smooth the images with a 6 pixel (3$\sigma$) Gaussian kernel, which smooths both sources into separate distinct objects. We use the brightest pixel of the primary source as an approximation as to the location of the AGN, and compute the location of the PSF asymmetry region relative to this point, avoiding contamination of the centroid from the events within the asymmetry region. 

The asymmetry region present in the Chandra PSF at small scales (see \cite{kashyap_2010}), which is thought to contain $\approx$5\% of the source counts within a 50$^{\circ}$ arc $0\farcs6-0\farcs8$\footnote{\url{https://cxc.cfa.harvard.edu/ciao/caveats/psf_artifact.html}}, is the most likely source of this anomalous point source. For each observation some of the events associated with the second source fall within the asymmetry region, and the 2 longest observations were taken at the same roll angle, resulting in overlapping asymmetry regions. The documentation suggests that around 5\% of the total counts associated with a source are found in this region, but at least for the 6-7 keV band 15\% of the counts are within the asymmetry region. 
We removed the asymmetry region from each event file and recombine them using \textit{merge\_obs}, and find the final uncontaminated point source location using \textit{wavdetect} We associate 6 counts with this region, and from our PSF simulations we would expect 3.5 counts for the same region at $\sim$1$\arcsec$ distance. While there is still an excess grouping of counts above the predicted level for the asymmetry region, given the uncertainties associated with this artefact we are unable to conclude that there is a statistically significant second source within the region. Figure \ref{fig:bubble_xray_hard_smooth} shows that there is a spatially coincident [O~{\sc iii}] excess at the same position. 

The black cross in Figure \ref{fig:bubble_xray_soft_smooth} shows final nuclear source location determined by this method (R.A.: 15:38:54.1575, decl +17:01:34.449). This source location is used to generate the simulated PSF using ChaRT and MARX, and as the central point for computing the surface brightness gradient. The brightest [O~{\sc iii}] emission is located at  R.A. 15:38:54.1379 decl +17:01:34.543 and is shown by the green cross in Figure \ref{fig:bubble_xray_soft_smooth}. The centroid of the near-IR emission, shown by the red cross in Figure \ref{fig:bulge}, is located at R.A 15:38:54.1551 decl +17:01:34.274. The distance between the nuclear source location and the brightest [O~{\sc iii}] point is 0$\farcs3 \pm 0 \farcs$2. There is a 0$\farcs 2\pm 0\farcs$1 separation between the near-IR centroid and the nuclear source location, and a 0$\farcs 4\pm 0 \farcs$1 separation between the near-IR centroid and the brightest [O~{\sc iii}] region. It is difficult to determine whether there is a significant discrepancy given the uncertainty in the nuclear source location (as detailed above), as well as the uncertainty in the brightest [O~{\sc iii}] point and the 0$\farcs$09 RMS residual offset of the astrometric match between the 2 observations. If the offset is real it gives us clues as to the geometry of the NLR in ways that are suggestive of selective extinction. The model used in \cite{Zhao2020} constrains the torus and inclination angles of the AGN as 32° and 26° respectively, suggesting the bi-cone could be aligned with the plane of the sky. If so, the Iron K$\alpha$ emission would be absorbed by the Compton-thick torus but [O~{\sc iii}] emission from the more distance narrow line region would still be observed. 

\subsection{Spectra and Morphology of the Bubble}

The soft emission coincident with the [O~{\sc iii}] bubble shows a broadly similar morphology to the [O~{\sc iii}] structure, with the same spatial extent. We see soft X-ray emission from the region NW of the nucleus containing as a dust lane, which may suggest that [O~{\sc iii}] emission is selectively obscured while soft X-ray emission is not. A deeper Chandra observation (a few hundred kiloseconds) would allow us to de-convolve the effects of the PSF and make better comparisons to the small-scale [O~{\sc iii}] structure. 

The best-fitting model for the soft (0.3-2 keV) emission we consider is the double APEC component, representing a multi-component thermally ionized plasma, with a reduced statistic indicative of a good fit. There is some correlation in the residuals, with an excess near 0.9 keV that can potentially be attributed to Ne IX emission, but may also have contributions from the [Fe~{\sc xviii}], [Fe~{\sc xix}], and [Ni~{\sc xix}] lines with similar energies. Figure \ref{fig:bubble_xray_neix} shows the Ne IX emission from the bubble, which in other systems is indicative of shocked emission  (e.g. NGC 3933 \citealt{Maksym2019}, NGC 4151 \citealt{wang2011deep}). The bulk of the soft emission is best-fitted by the APEC model with a temperature of 0.80 keV, with the other component modelling the very soft 0.5 keV excess at a poorly constrained temperature of $\approx$0.1 keV. 

When we consider a single CLOUDY model, with no role for a thermal component, we get a best-fitting model with a reduced statistic indicative of a strong fit, but with residuals indicating that the model does not explain all the spectral features, including the very soft excess. Introducing a second CLOUDY component as a potential alternative model for this excess does not improve the fit, with one component completely dominating over the other. Fitting a single or double component to a region covering around 2.5 kpc in diameter means averaging over a potentially complex system which is unlikely to be well parameterised with a single column density or ionization parameter. The column density we find is at the lower limit allowed by the CLOUDY model (log N$_{\textrm{H}}$=19.0 cm$^{-2}$), suggesting a very low column density an order of magnitude below the galactic column density. We were unable to compute a confidence interval for either parameter, so the upper limit is unconstrained. The medium could be clumpy, with knots of high column density, but the best-fitting column density is likely an underestimate of the true average column density across this region.

The surface brightness profiles indicate that there is extended soft (0.5-1.5 keV) X-ray emission out to 2$\farcs$5$\arcsec$(1500 pc), which is preferentially extended along the long axis of the [O~{\sc iii}] bubble. From the spectral fitting and the BPT mapping we see that emission observed is consistent with a photo-ionization an AGN-like source. One comparison that can be made is to Extended Narrow Line Regions (ENLRS) in other systems, which display the same signatures of AGN-influence across kpc scales. X-ray studies of sub kpc Extended Narrow Line Regions (ENLRS)  \citep[e.g.][]{Paggi2012,bogdan2017probing,fabbiano2018deep} have shown that the gas populations within ENLRs are multi-phase, with roles for both photo-ionized and collisionally ionized plasma. Fitting the thermal and photo-ionization models both models simultaneously also produces a strong fit, with the photo-ionization component dominating and a similar low temperature thermal component for the soft excess. Contributions from both components may therefore be needed to fully explain the ENLR emission. 

\subsection{Pressure and Density}
\label{sec:pressure_density}
The reduced statistic indicates that the soft emission is well-fitted by the double APEC model which represents a multi-component thermally-ionized plasma. If we assume the gas is dominated by a collisional outflow we can estimate physical parameters of the gas within the bubble. From the best-fitting emission measures (normalisation) and temperatures of the double APEC model we can derive an average thermal pressure by this equation from \cite{Paggi2012}:
\begin{equation}
EM = \frac{10^{-14}}{4\pi D_A^2 (1+z)^2} \int n_e n_H dV
\end{equation} 
where $D_A$ is the angular distance and z is the redshift. We assume an ideal proton-electron gas (i.e n$_e$=n$_\textrm{H}$) and the pressure-temperature relation of $p_{\textrm{th}}$ = $2n_ekT$. We also assume the density and temperature of the gas is approximately constant across the bubble. 

We measure the on-sky size of the bubble from the [O~{\sc iii}] HST image, assuming a similar size for the coincident X-ray gas population. As is done in \cite{Maksym2019}, we model the shape of the bubble as an ellipsoid, with the z-axis size estimated as the shorter axis of the ellipsoid. We calculate all spatially dependent parameters (e.g. \textit{L$_{\textrm{kin}}$}, \textit{t$_{\textrm{cross}}$}) relative to the longest axis of the ellipsoid. The major semi-axes of the bubble are estimated as 1$\farcs$40 $\times$ 0$\farcs$95 $\times$ 0$\farcs$95 ($\pm 10\%$) which is equivalent to (850$\pm$90 pc $\times$ 590$\pm$60 pc $\times$ 590 $\pm$ 60 pc). This gives a volume of 1.2 $\pm$ 0.2 kpc$^3$. For X-ray emitting plasma we expect a density of around 1 cm$^{-3}$ and thermal pressure of p$_{\textrm{th}}$ $\approx 10^{-10}$ dyne cm$^{-2}$ \citep{Paggi2012}.

\begin{table}
\centering
\renewcommand{\arraystretch}{1.5} 

\begin{tabular}{lll}
\textbf{Parameter}                     & \textbf{APEC$_1$ Model}               & \textbf{APEC$_2$ Model}                   \\ 
\hline
kT (keV)                               & 0.80$^{+0.09}_{-0.11}$                         & 0.10$^{+0.06}_{-0.08}$                    \\
EM                                     & 5.0$^{+1.8}_{-1.4}\ \times \ 10^{-6}$ & 4.2$^{+\textrm{u}}_{-3.7}\ \times 10^{-4}$\      \\
APEC Flux (erg cm$^{-2}$ s$^{-1}$)                     & 1.5$\pm0.6 \ \times \ 10^{-14}$ & 1.6$\pm1 \ \times \ 10^{-13}$      \\
APEC Luminosity (erg s$^{-1}$)               & 2.9$\pm1.2 \ \times \ 10^{40}$ & 3.1$ \pm 2.1 \ \times \ 10^{41}$      \\

Electron Density (cm$^{-3}$)           & 0.17$\pm$0.04                         & 1.6$^{+\textrm{u}}_{-0.7}$                       \\
Pressure (dyn cm$^{-2}$)               & 4.2$^{+1.2}_{-1.1} \times$ 10$^{-10}$         & 4.8$^{+\textrm{u}}_{-4.4}$ $\times 10^{-10}$          \\
Total Energy Budget (erg)              & 3.8$^{+1.2}_{-1.2}$~$\times$ 10$^{55}$        & 4.3$^{+\textrm{u}}_{-4}$ $\times 10^{55}$         \\
Shock Velocity (km s$^{-1}$)           & 780$^{+50}_{-50}$                            & 270$^{+80}_{-110}$                        \\ 
Sound Speed (km s$^{-1}$)                    & 450$^{+30}_{-30}$                 & 160$^{+50}_{-60}$                             \\
Cooling Time (Myr)                    & 42$^{+14}_{-13}$                 & 4.4$^{+\textrm{u}}_{-4}$                             \\

Crossing Time (Myr)                    & 2.1 $\pm$0.3                          & 6.1$^{+1.9}_{-2.5}$                        \\
Kinetic Luminosity (erg s$^{-1}$)      & $\geq$5.7$\pm$1.9 $\times$10$^{41}$   & $\geq$2.2$^{+\textrm{u}}_{-2}\times$10$^{41}$  \\
L$_{\textrm{kin}}$/ L$_{\textrm{bol}}$ & $\geq$0.8$^{+0.3}_{-0.2}$\%           & $\geq$0.30$^{+\textrm{u}}_{-0.2}$\%              \\ 
           
\end{tabular}

\caption{The results of the calculations based on the best-fitting APEC models detailed in Section \ref{sec:pressure_density}. Units are given next to the parameter name where appropriate. The upper and lower bound or 1$\sigma$ uncertainty are given where they have been computed. }
\label{tab:apec_results}
\end{table}

\subsection{Feedback Effects from X-ray Spectroscopy}
\label{sec:apec_calcs}
We follow the method used in \cite{Paggi2012}, \cite{Sartori2016} and \cite{Maksym2019} to estimate the interaction between the AGN and the ISM within the bubble. The results of the following calculations are shown in Table \ref{tab:apec_results}. Many of the equations and principles are adapted from the study of bubbles/cavities within dense galaxy clusters \citep[e.g.][]{birzan2004systematic,2012NJPh...14e5023M,2014MNRAS.442.3192V}.
Using the estimated volume and $p_{\textrm{th}}$ we can calculate an average thermal energy budget using $E_{\textrm{th}} = \frac{\gamma}{\gamma-1}p_{\textrm{th}}V$. the heat capacity ratio, $\gamma$, of a non-relativistic monoatomic gas is 5/3. We calculate a shock velocity for the gas using:
\begin{equation}
    v_{\textrm{shock}} \approx 100 \ \textrm{km s}^{-1} \times (kT/0.013 \ \textrm{keV})^{0.5}
\end{equation}
per \cite{Raga_2002}. The crossing time is then simply the length of the relevant axis of the ellipsoid divided by the shock velocity.

We can calculate a sound speed c$_{\textrm{s}}$ from \begin{equation}
     c _{\textrm{s}}= (\gamma kT/\mu m_H)^{0.5}
\end{equation} \citep{birzan2004systematic} where $\mu$=0.62, and $\gamma$=5/3. $\mu$ is a scaling factor for the average particle mass taken from \cite{birzan2004systematic}. 

We use the best-fitting APEC models to calculate an unabsorbed and unconvolved energy flux for each component using the Sherpa tool \textit{calc\_energy\_flux}, the results of which is shown in Table \ref{tab:apec_results}. The corresponding thermal luminosities are calculated by integrating the flux over 4$\pi$ steradians at the luminosity distance of \textit{d}=133$\pm$9 Mpc. The cooling time is then simply the total thermal energy budget divided by this luminosity. For the hotter component (APEC$_1$), the cooling time is 20-30x longer than the crossing time, suggesting no extra heating source is required. For the cooler component, the crossing time is longer than the cooling time, suggesting an extra heating component may be needed, but both of these time estimates are not well-constrained. 

We can calculate a lower limit on the fraction of kinetic power heating the ISM as L$_\textrm{K}$ $\geq $ $E_{\textrm{th}}$/t$_{\textrm{cross}}$ \citep{Paggi2012}. We can then compare these to our calculated bolometric luminosity to estimate the efficiency of the feedback processes. 

The crossing time of the bubble for either component is much longer than the light travel time (about 5 kyr), suggesting the bubble is much older than the current period of variability and not related to a recent shock or outflow. The ratio of kinetic to bolometric luminosity of 0.8$^{+0.3}_{-0.2}$\% from the dominant APEC component is indicative that there is efficient feedback that could drive hot gas outflow and clear out the gas over large time-scales, as \cite{10.1111/j.1365-2966.2009.15643.x} suggest a lower threshold of L$_{\textrm{kin}}/L_{\textrm{bol}}$ > 0.5\% for a two-stage feedback model. There is no evidence of feedback efficient enough to directly entrain the ISM.

Given the decrease in L$_{\textrm{bol}}$ on time-scales much shorter than both the crossing time and the cooling time of either of our gas components, our estimate of the kinetic luminosity may be based on emission from gas which has not had time to respond to the decreasing bolometric luminosity. This may lead to overestimates of the "instantaneous" feedback efficiency in AGN which show large luminosity variations.

Observations with better angular resolution and effective area with a proposed mission like Lynx \citep{gaskin2018lynx} are required to better resolve the full effects of feedback within the bubble. 

\subsection{Venting}
One possibility is that the [O~{\sc iii}] bubble and associated extended soft X-ray emission is a signature of hot gas ablated from the inner disk venting through a gap in the torus. The theory is that an outflow directed into the plane of the galaxy could encounter dense molecular clouds, which may be shocked but not efficiently entrained by the outflow. The outflow could then continue to move along the most efficient path, likely to be out of the plane of the galaxy away from high-density obstacles, similar to a vent \citep{2018MNRAS.476...80M}. Such vented outflows may carry entrained gas, or even shock lower-density material outside the plane, and those shocks could produce X-rays and [O~{\sc iii}] gas such as we see in NGC 5972. A comparable example may be found in IC 5063, described in \cite{2021ApJ...917...85M}. IC 5063 demonstrates arc-second ($\approx$700 pc) scale forbidden emission lines loops which have been predicted by successful models of jet-ISM interactions \citep[e.g.][]{2018MNRAS.476...80M}. The loop in IC 5063 is perpendicular to the bi-cone, which we also see in NGC 5972. We see preferential extent of the [O~{\sc iii}] bubble in the E-W direction (and soft X-ray emission in the E) and we know that the bi-cone is aligned approximately N-S in line with the EELRs. 
\cite{2021ApJ...917...85M} reliably identify the loop in IC 5063 only in HST [S~{\sc ii}] observations, although there is also some evidence in [O~{\sc iii}], H$\alpha$ and H$\beta$. Via emission line ratios they identify the loop as consistent with LINER emission, whereas our  emission is consistent with a Seyfert-like source. As \cite{Keel2012} finds for HV, we find the implied age of the bubble to be longer than the current variability, suggesting it may not have formed as a result of an AGN state change. It could instead be a signature of venting, which could occur over a longer time-scale. 

\subsection{Spectra and Morphology of EELRs}

The soft X-ray spectra of the north EELR are best-fitted by the CLOUDY model, meaning that as we expect the gas is consistent with ionization by AGN emission. The best-fitting column density of 22.7$\pm$0.2 cm$^{-2}$ and ionization parameter U=1.2$\pm$0.3 suggest we are observing a moderately obscured and highly photo-ionized gas. Given the inferred geometry of the system \citep[e.g. in][]{Keel2015,Zhao2020}, the EELRs are located within the bi-cone of the AGN with a large opening angle into the plane of the sky and directly photo-ionized by the AGN emission. The source of the obscuration is not certain, but the X-ray gas may be screened by the large amount of [O~{\sc iii}] gas, particularly at larger projected distances. There is also a potential role for dust obscuration given that \cite{Keel2017} observe a dust lane to the NE of the bubble.

We have had to average the model over the entire EELR, but it is possible that there are large variations in column density and ionization across the EELR given the large variation in projected extent from the AGN. In Figure \ref{fig:arms_oiii_xray_smooth} the soft X-ray emission from the North EELR shows a similar surface brightness over a large range of projected extent from the EELR ($\approx$ 15$\arcsec$ or 9 kpc). Irrespective of the recent variability of the AGN, if the column density was constant across this region we would expect a non-uniform intrinsic X-ray luminosity as we don't observe a brightness gradient. For a uniform point source we would expect a SB $\propto$ 1/r$^2$ gradient in brightness with increasing distance from the AGN.

The $3\sigma$ smoothed soft X-ray emission in Figure \ref{fig:arms_oiii_xray_smooth} appears somewhat diffuse, potentially as a consequence of smoothing relatively few counts across a larger area. However at low significance the 1$\sigma$ adaptively smoothed soft X-ray image shows region of higher emission, potentially correlated with the [O~{\sc iii}] EELR emission. The [O~{\sc iii}] emission is clumpy, with knots of high surface brightness emission, so if the X-ray emitting gas is co-spatial it is possible that it is also somewhat clumpy, with variable column density and unresolved knots and cavities. It is generally coincident with the [O~{\sc iii}] gas, potentially suggesting a  multi-phase medium, however there are regions with soft X-ray emission which display no co-spatial [O~{\sc iii}] emission (e.g. isolated contour to NW of North EELR, extended contour to the E of the North EELR). In the south EELR the soft X-ray appears to follow the curved [O~{\sc iii}] tail but with the X-ray peak at a larger projected distance, which is suggestive of a weak shock inducing soft X-ray emission. The isolated contour west of the North EELR is aligned with the jet orientation inferred from the VLASS observation, and so it may be unrelated to the EELR but a signature of a jet-induced shock producing soft X-rays. 

\cite{Fabbiano2019} have shown that soft X-ray emission can be a useful diagnostic of X-ray emission over longer time-scales than [O~{\sc iii}] emission (on the order 10$^7$ yr). A deeper Chandra observation would allow us to constrain the AGN variability over Myr time-scales. This would allow us to average over individual duty cycles and probe the overall accretion rate of the AGN. Given the spacing of the radio lobes (330 kpc), shown in Figure \ref{fig:radio_lobes}, and the current (67$^{\circ}$ misalignment of the jets relative to the lobes is suggestive that there has been significant jet activity and variation on similar time-scales. The radio jet within the north EELR is approximately 15$\arcsec$ from the nuclear source, which corresponds to about 9 kpc, without considering projection effects which would increase this distance. To calculate an order of magnitude minimum jet age we assume a jet velocity of anywhere between 0.05-0.9c \citep{livio1999astrophysical}, gives a minimum jet age of 30-600kyr. This estimate is the same order of magnitude as the current observed variability, suggesting there may be a relationship between the jet and the AGN state change.

The long axis of the [O~{\sc iii}] bubble is also misaligned with the other features. Similar misalignments between jets and lobes are seen in "re-started" AGN", where the jets undergo a dramatic shift in orientation, but the radio lobes relics from the previous emission are still observed \citep[e.g.][]{10.1093/mnras/stt1606,2010ApJ...717L..37H}. We see no evidence for a radio lobe associated with the new jet.
The jet shift may be due to significant interaction or merger, which we do see other evidence for in this galaxy, or evidence of a binary black hole \citep[e.g.][]{2002Sci...297.1310M,2007ApJ...659L...5C}. We speculate that the overall misalignment could be explained by the presence of a binary SMBH system, which could support a more complex system of outflows, jets and ionization cones \citep[e.g.][]{2017MNRAS.465.4772R}. Such double SMBH have been observed \citep[e.g.][]{2003ApJ...582L..15K,2004ApJ...604L..33Z, koss2016nustar, 2022A&A...658A.152V}, but if both are actively accreting (dual AGN) they are often associated with double-peaked emission lines that are not resolved here \citep{shen2010identifying}. The existence of double-peaked lines would also depend on the dynamics of the binary system. Double-peaked lines can also sometimes be caused by the rapid rotation of the BLR without the presence of a binary \citep{eracleous2009double}. A dual AGN could also explain our difficulty in constraining the centroid because two active SMBHs emitting hard nuclear spectra could appear as 2 almost-unresolved sources given the sub-arcsecond resolution of the ACIS detector. Higher resolution radio observations to better resolve the bubble and jets would provide greater insight into the cause of the misalignment.

The near-IR emission from NGC 5972 (Figure \ref{fig:bulge}), is consistent with a typical galaxy bulge, and we do not see any evidence of structure that could suggest a resolved binary SMBH system. Given the 0$\farcs$15 PSF of this observation we can set an upper limit for the separation of a theoretical binary  of $\approx$0\farcs2, which corresponds to $\approx$100 pc at the redshift of NGC 5972. We cannot constrain the possibility of a binary SMBH with a separation between $\approx$10-100 pc.

Spectroscopy of extended soft X-ray emission in other AGN tend to be best-fitted by some combination of photo-ionization and thermal components \citep[e.g.][]{wang2011deep,Paggi2012,fabbiano2018deep}. In our case we do not find that the introduction of a thermal component improves the fit, and indeed we find that the contribution of the component is minimised by the fitting algorithm. 
The poor fitting of any model to the south arm is not necessarily suggestive that it is not well-fitted by a photo-ionized or thermal model. The south arm region contains fewer counts than the north region. Averaging over a large-scale, complex system with low X-ray surface brightness means our spectra are poorly resolved and may not be well-approximated by our simple models. 

\subsection{[O~{\sc iii}] to Soft X-ray Ratios}

\cite{2006A&A...448..499B} showed that the ratio of [O~{\sc iii}] to soft X-ray (0.5-2 keV) flux is inversely proportional to the ionization parameter U if both the X-ray and [O~{\sc iii}] emission are produced by the same photo-ionized medium. We can make comparisons to ratios observed in NGC 4151 and Hanny's Voorwerp. For clouds within NGC 4151 \cite{wang2009highest,wang2011deep} find ratios of $\sim$10-15, independent of the cloud distance from nucleus. They also find outliers with ratios as high as 100, implying lower ionization parameters, which they suggest may be due to intervening absorption. A constant ratio implies a constant ionization parameter, which requires a density $\propto r^{-2}$ as expected for the NLR or a nuclear wind.

For Hanny's Voorwerp \cite{Fabbiano2019} find a [O~{\sc iii}] to soft X-ray ratio of $\sim$200, which is considerably higher than our values for either EELR. This suggests that NGC 5972 is more highly ionized, which is backed up by the ionization parameter of U=1.2 inferred from the spectral fitting. 

For the [O~{\sc iii}] bubble, where we find a ratio of $\sim$3, this is comparable to the ratios \cite{wang2011deep} find within the radio jets. This high ratio may indicate very efficient X-ray photo-ionization, or that some [O~{\sc iii}] flux is obscured. 

Analysis of the ratio maps are limited by the low X-ray surface brightness and the adaptive smoothing, but we do find that in the EELR the highest ratios are found in the gas nearest to the AGN, suggesting that the regions of high [O~{\sc iii}] flux have lower soft X-ray emission, despite the proximity to the AGN. One hypothesis is that this [O~{\sc iii}] emission comes from dense regions which screen the X-ray emitting gas, preventing photo-ionization. Alternatively the soft X-ray emission is absorbed by [O~{\sc iii}] gas in the line of sight in these dense regions. Deeper X-ray observations would allow comparison of this ratio on smaller spatial scales with a useful signal to noise ratio. 

\subsection{STIS Kinematic Profiles}
\label{sec:stis_kin_discuss}
The literature value of the redshift calculated from 21cm observations of neutral hydrogen is z=0.02964 \citep{2014MNRAS.440..696A}, which we use to remove the wavelength shift introduced by the Hubble flow and probe only the kinematics of the gas within the galaxy.
Using the literature value for calibrations of gas velocities via Doppler shift of the spectral lines relative to their vacuum wavelengths produces a turnover in velocity (positive to negative in the LOS) which is hard to explain as a signature of an outflow. By using this redshift we are inherently assuming that the neutral hydrogen is at the same redshift as the excited gas and stellar components, which is not necessarily a good assumption in a complex system.  
To infer a better redshift estimate we turn to observations of the Calcium (Ca II) absorption triplet at vacuum wavelengths 8498, 8542, 8662 \AA, which is a tracer of old stellar populations \citep{andretta2005ii}. We choose this feature because it is in the NIR, so it is easy to calibrate and is less affected by reddening and dust absorption. We review spectra of NGC 5972 from the Sloan Digital Sky Survey (SDSS) DR16 \citep{2020ApJS..249....3A}, with a spatial single fibre resolution of 3$\arcsec$ marginally sufficient to resolve the central region. In the spectra the Calcium triplet is clearly visible, with observed wavelengths 8752, 8798 and 8923 \AA \ respectively, which corresponds to an average redshift of 0.0300. This redshift change is a velocity shift on the order of 100  km s$^{-1}$. Using this redshift for our velocity calibration instead we no longer measure a turnover in the velocity. 

The G430L spectra showed a systematic wavelength offset from the G750M spectra based upon comparison of the redshift of the H$\alpha$ and H$\beta$ lines, which we would expect to be emitted by the same gas population. The source of this systematic shift is unclear as it appears in both the spectra produced by the Hubble Legacy Archive's legacy pipeline as well as our reprocessed spectra. This shift has the effect of a 2.5\AA \ (150  km s$^{-1}$ at the wavelength of the [O~{\sc iii}] line) blueshift, which we correct for in our [O~{\sc iii}] velocity profiles. The radial velocity of the [O~{\sc iii}] gas peaks at $\sim$300  km s$^{-1}$, which may be strong enough to drive shocks within the lower temperature X-ray emitting medium suggested by the second APEC component. However this component only weakly contributes to the observed spectra and is not well-constrained. The implied radial velocity of the [O~{\sc iii}] gas is not strong enough to drive shocks within the main thermal component (APEC$_1$) unless it were only a component of the true velocity of the outflow due to LOS projection effects. Given our calculated sound speeds the 780 km s$^{-1}$ outflow we estimate would be strong enough to drive fast shocks capable of photoionising the precursor material in the plane of the sky. Assuming a bi-conical outflow as is done in \cite{Fischer2013}, our spectra may not be deep enough to measure the redshifted [O~{\sc iii}] which could be heavily obscured. Alternatively the bubble might contain a multi-phase medium, with distinct [O~{\sc iii}], H$\alpha$ and soft X-ray emitting populations which weakly collimate each other. This could also explain the large disparity in gas densities between the APEC and [S~{\sc ii}] calculations.

In the HST images of the [O~{\sc iii}] bubble, particularly Figure \ref{fig:stis_loc}, we observe a number of small scale structures within the bubble itself, indicated by regions of varying surface brightness. These may be indicative of smaller bubble or loop like filaments embedded within the main structure. The STIS slit location passes through several of these features, which may be complicating our observed velocity profile across the bubble. These bubble-like structures may be indicative of outflow variations. There might also be variable extinction across the bubble, causing small-scale [O~{\sc iii}] reddening. 
\cite{Fischer2013} discusses models to constrain opening angle and orientation based on observed velocity profiles of NLR gas. Our velocity profile is more complex than predicted by their models, but they do suggest asymmetries in the velocity profile can sometimes be explained by line-of-sight sampling effects. 

\subsection{Kinematic Comparisons}
\label{sec:kin_compar}
\cite{Keel2015,Keel2017} and Finlez et al. 2022 (in preparation) have both analysed the kinematics of the central region of NGC 5972 using 2D velocity maps derived from Integral Field Unit (IFU) spectroscopy and Fabry–Perot interferometric (FPI) data. We can make comparisons between their work and the velocity profiles we derive from the STIS slit spectroscopy. Overall we find good agreement with the kinematics \cite{Keel2017} observe, given the differences in resolution between the ground-based IFU data and the Hubble STIS observations.

The Gemini Multiple-Object Spectrometer (GMOS) data presented by \citeauthor{Keel2017} (\citeyear{Keel2017}; Fig. 12), shows the $\approx$100 km s$^{-1}$ blueshift of the H$\alpha$ line emission in the nucleus. The increase in velocity dispersion west of the nucleus is seen in the GMOS map ($\approx$ 100 km s$^{-1}$) as well as the FPI data (130 km s$^{-1}$ in [O~{\sc iii}]5008\AA). This is consistent with the velocities we observe given the beam smearing effect which reduces the observed velocities for the lower spatial resolution of the ground-based observations.

Finlez et al. 2022 (in preparation) have used the Multi Unit Spectroscopic Explorer (MUSE) instrument on the Very Large Telescope (VLT) to do wide-field integral field spectroscopy of NGC 5972. Our results are broadly consistency with the velocity maps they observe. Within the central region they see evidence of outflows up to 300 km s $^{-1}$ but are limited by the instrumental spectral resolution of MUSE. We independently observe similar outflow velocities, suggesting that the 300 km s$^{-1}$ outflows may be a real feature.

\section{Conclusions}

In this paper we present new Chandra ACIS X-ray observations of NGC 5972 alongside new HST STIS spectroscopy. NGC 5972 is one of the 19 "Voorwerpjes" galaxies, with an Extended Emission Line Region that has been used as a quasar light echo to infer a 2.1dex decrease in bolometric luminosity over the last 50 kyr. 
We detect broadband X-ray emission from NGC 5972 coincident with an arc-second scale bubble-like structure seen in HST [O~{\sc iii}] imaging which contains the central engine. Simulations of the Chandra PSF show that the soft emission (0.5-1.5 keV) is spatially extended out to 2$\farcs$5 in the same direction as the bubble. Spectral modelling shows that the soft emission is compatible with a combination of photo-ionized and collisionally ionized gas. Our best model suggests two populations of collisionally ionized gas with temperatures of $\sim$0.80 keV and $\sim$0.1 keV. From this model we have been able to estimate other physical parameters, including the effects of feedback on the ISM. We calculate an age of the bubble of at least $\sim$2 Myr, much longer than the current variability time-scale ($\sim$ 50 kyr). We find evidence of efficient feedback, with ratios of kinetic to bolometric luminosity of $\sim$1\%, indicative of feedback strong enough to drive out the gas on long time-scales under a two-stage feedback model. We discuss the possibility that this ratio is overestimated given the known decrease in the bolometric luminosity on time-scales much shorter than the crossing time of the bubble.

The STIS spectroscopy has allowed us to probe the kinematics of the gas across the bubble. We find evidence of outflow velocities up to 300  km s$^{-1}$ from the [O~{\sc iii}] line. Given our estimates of sound speeds, this velocity would only be directly strong enough to drive shocks in the lower temperature population. However we suggest that the measured radial velocity may only be a component of the $780$ km s$^{-1}$ outflow implied by the higher temperature gas, given a small LOS projection effect. An outflow of this velocity could drive strong shocks in the plane of the sky \citep{1996ApJS..102..161D}. The overall unsymmetrical velocity profile is complex and not well-explained by a simple outflow model, which may partly be explained by a line of sight projection effect.

NGC 5972 shows misalignment on multiple spatial scales, including the jets, lobes and EELRs, which is not well-explained by the traditional simplistic AGN model. The current radio jets, seen in VLASS data, are not aligned with the radio lobes, which is similar to the misalignment we see in "restarted" radio AGN. Like a majority of the "Voorwerpjes" galaxies, NGC 5972 shows evidence, such as tidal tails, of a recent merger or significant interaction. On a smaller scale the long axis of the [O~{\sc iii}] and X-ray structure is perpendicular to the bi-cone, which is aligned with the EELRs. These results might imply the presence of a double SMBH, which could support a more complex system of outflows, jets and lobes. Using NIR AO imaging we constrain the upper limit for the separation of a possible binary system as $\approx$100 pc. 

We detect soft X-ray emission from diffuse hot gas coincident with the [O~{\sc iii}] EELRs. Ratios of [O~{\sc iii}] to soft X-ray flux are comparable to those seen in other EELRs. Spectral analysis of these EELRs is consistent with a photo-ionising source, but deeper X-ray observations are required to constrain the model parameters and make better morphological comparison to the [O~{\sc iii}] emission. 

Much of our analysis is limited by the depth of the X-ray observations. With a longer Chandra observation of a few hundred kiloseconds, broken into fewer sections with different roll angles we would be able to simultaneously probe the extended soft emission from the EELR as well as de-convolve the hard X-ray emission from the nucleus in order to confirm the nuclear source centroid and extent with better accuracy. With deeper Chandra observation of the EELRs it would be possible to make more granular comparisons to the [O~{\sc iii}] structure. It may be possible to estimate the historic variability on Myr time-scales using the soft X-ray emission from the EELRS as a light echo as has been done for Hanny's Voorwerp. Probing longer time-scales than is possible using optical EELR lets us average over multiple duty cycles, and probe the time-scales relevant to to the formation of extended jets and radio lobes. 

\label{sec:conclusions}

\section*{Acknowledgements}
T.H. would like to thank the University of Southampton (Astrophysics with a Year Abroad programme) and the Center for Astrophysics | Harvard \& Smithsonian, where this research was carried out.
W.P.M acknowledges support by Chandra grants GO8-19096X, GO5-16101X, GO7-18112X, GO8-19099X, and Hubble grant HST-GO-15350.001-A.
V.N.B. gratefully acknowledges assistance from National Science Foundation (NSF) Research at Undergraduate Institutions (RUI) grants  AST-1909297. Note that findings and conclusions do not necessarily represent views of the NSF.
A.M. was supported by  the SAO RAS government contract approved by the Ministry of Science and Higher Education of the Russian Federation.
L.F.S. acknowledges the financial support of the Swiss National Science Foundation.

This research is based on observations made with the NASA/ESA Hubble Space Telescope obtained from
the Space Telescope Science Institute, which is operated by the Association of Universities for
Research in Astronomy, Inc., under NASA contract NAS 5-26555.

\section*{Data Availability}

The raw data products used in this analysis are publicly available on the Chandra Data Archive (\url{https://cxc.cfa.harvard.edu/cda/}) and the Hubble Legacy Archive (\url{https://hla.stsci.edu/}).
Data products produced for this paper will be provided on reasonable request to the authors.



\bibliographystyle{mnras}
\bibliography{main} 








\bsp	
\label{lastpage}
\end{document}